\documentclass[aps,preprint,floatfix,nofootinbib,showpacs]{revtex4-1}
\pdfoutput=1
\usepackage{graphicx,color}
\usepackage{hyperref}
\usepackage{placeins}

\usepackage{amsmath}
\usepackage{amsfonts}
\usepackage{amssymb}
\def\ga{\mathrel{\raise.3ex\hbox{$>$\kern-.75em\lower1ex\hbox{$\sim$}}}}
\def\la{\mathrel{\raise.3ex\hbox{$<$\kern-.75em\lower1ex\hbox{$\sim$}}}}

\newcommand{\lsim}{\raisebox{-0.13cm}{~\shortstack{$<$ \\[-0.07cm] $\sim$}}~}
\newcommand{\gsim}{\raisebox{-0.13cm}{~\shortstack{$>$ \\[-0.07cm] $\sim$}}~}

\begin{document}

{\small
\begin{flushright}
CNU-HEP-15-06
\end{flushright} }

\title{
Enhanced Charged Higgs Production through 
$W$-Higgs Fusion \\ in $W$-$b$ Scattering}

\def\thefootnote{\fnsymbol{footnote}}

\def\slash#1{#1\!\!/}

\medskip

\author{
Abdesslam Arhrib$^{1,2}$,
Kingman Cheung$^{2,3,4}$, 
Jae Sik Lee$^{2,5}$, and
Chih-Ting Lu$^3$
}
\affiliation{
$^1$ D\'epartement de Math\'ematiques,
Facult\'e des Sciences et Techniques,
B.P 416 Tangier, Morocco \\
$^2$ Physics Division, National Center for Theoretical Sciences,
Hsinchu, Taiwan \\
$^3$ Department of Physics, National Tsing Hua University,
Hsinchu 300, Taiwan \\
$^4$ Division of Quantum Phases and Devices, School of Physics,
Konkuk University, Seoul 143-701, Republic of Korea \\
$^5$ Department of Physics, Chonnam National University, \\
300 Yongbong-dong, Buk-gu, Gwangju, 500-757, Republic of Korea
}
\date{\today}

\begin{abstract}
We study the associated production of a charged Higgs boson
with a bottom quark and a light quark at the LHC via
$ p p \to H^\pm\,b\,j$ in the Two Higgs Doublet Models (2HDMs). 
Using the effective $W$ approximation, we show 
that there is exact cancellation among various Feynman diagrams 
in high energy limit.
This may imply that the production of charged Higgs can be 
significantly enhanced in the presence of
large mass differences among the neutral Higgs bosons
via $W^\pm$-Higgs fusion in the $ p p \to H^\pm\,b\,j$ process.
Particularly, we emphasize the potential enhancement due to a light
pseudoscalar boson $A$, which is still allowed by the current data by
which we explicitly calculate the allowed regions in $(M_A,\,\tan\beta)$
plane, and show that the production cross section can be as large as 0.1 pb
for large $\tan\beta$.
We also show that the transverse momentum distribution of the $b$ quark
can potentially distinguish the $W^\pm$-$A$ fusion diagram
from the top diagram. 
Finally, we point out further enhancement when we go beyond the 2HDMs.
\end{abstract}

\maketitle

\section{Introduction}
%
A new scalar boson $h$ was  discovered 
in the run I of LHC with $7\oplus 8$ TeV energies in 2012 
\cite{cmsdiscovery,atlasdiscovery}. 
The combined measurement of the mass of the boson 
performed by the ATLAS and CMS  collaborations based on the data from 
$h\to \gamma \gamma$ and $h\to ZZ\to 4l$ channels is 
$m_{h} =$ 125.09 $\pm$ 0.21 (stat.) $\pm$ 0.11 (syst.) GeV~\cite{Aad:2015zhl}.
Furthermore, the measured properties of the  new particle
are best described by the standard-model (SM) 
Higgs boson \cite{Cheung:2013kla,update}.

The mission of the new LHC run at 13 TeV (and later upgraded to 14 TeV) 
is two folds: the first task is
the improvement of the scalar boson mass and scalar boson coupling
measurements and the second one
would be to find a clear hint of new physics.
By performing accurate measurements of the scalar boson
couplings to the SM particles would be helpful to determine
if the Higgs-like particle is indeed the SM Higgs boson
or a Higgs boson that belongs to a higher representation,
such as models with extra Higgs doublets, extra triplets, or singlets.
Most of higher Higgs representations with extra doublet or triplet Higgs 
fields predict in their spectrum one or more singly- or doubly-charged 
Higgs bosons.
A discovery of such charged Higgs bosons
would be an indisputable signal of new physics.

In the two-Higgs-doublet models (2HDMs) or the minimal supersymmetric standard
model (MSSM), the charged Higgs boson 
can be abundantly produced both at hadron and 
 $e^+e^-$ colliders. At hadron colliders, the charged Higgs boson
can be produced through several channels:
\begin{itemize}
\item Production from top decay. If the mass of
the charged Higgs boson is  smaller than $m_t -m_b$, 
the production of $t\bar{t}$ pairs provides an excellent
source of the charged Higgs bosons. If kinematically allowed, 
one of the top and anti-top quarks, say the anti-top quark 
can decay into $H^- \bar b$, 
competing with the SM decay of $\bar t \to W^- \bar b$. 
This mechanism $pp\to t\bar{t}\to t\bar{b}H^-$ can provide an 
important source of light charged Higgs bosons
and offers a much cleaner signature than that of direct production.

\item Single charged Higgs production.  The most important ones are 
$gb \to tH^-$ and $gg\to t\bar{b}H^-$ \cite{single}. These are QCD processes,
and thus the cross sections are expected to be large. 
We can also have a single charged Higgs boson produced in association with 
a $W^\pm$ gauge boson via
the loop process $gg \to W^\pm H^\mp $ or the tree level process
$b\bar{b} \to  W^\pm H^\mp $ \cite{wh}.
Similarly, the single charged Higgs boson can be produced in association with 
 a Higgs boson: $q\bar q' \to W^{\pm *} \to \phi H^\pm$ where 
$\phi$ denotes one the the three neutral MSSM Higgs bosons~\cite{cpyuan}.
Most of these 
processes are of the Drell-Yan type, they are 
expected to give substantial cross sections only for the charged Higgs 
mass below about 200 GeV.

\item Single charged Higgs boson production
associated with a bottom quark and a light
quark $qb\to q' H^+ b$ in the MSSM framework in which the 
neutral heavier Higgs bosons are almost degenerate~\cite{moretti}. 

\item Charged Higgs pair production through $q\bar{q}$ 
annihilation \cite{hadronic} or gluon fusion.

\item Resonant charged  Higgs production  $c \bar s\to H^+$, $c \bar b\to H^+$ 
\cite{Dittmaier:2007uw}.
\end{itemize}

At the Tevatron and LHC, detection of light charged Higgs boson with
$M_{H^\pm} < m_t - m_b$ is straightforward from $t\bar{t}$ production 
followed by the decay $\bar{t}\to \bar{b}H^-$ or $t\to bH^+$.
Such a light charged Higgs boson 
can be detected for any value of $\tan\beta$ in
the $\tau\nu$ decay which is indeed the dominant decay mode. 
The ATLAS and CMS have already had an exclusion on $B(t\to bH^+) \times
B(H^\pm \to \tau \nu)$ based on this decay channel 
\cite{Aad:2014kga,Khachatryan:2015qxa}.

In the MSSM and 2HDMs, the heavy charged Higgs boson with $M_{H^\pm}\ga m_t$
would decay predominantly into $t\bar{b}$.
The experimental search is rather difficult due to large irreducible 
and reducible backgrounds associated with $H^+\to t\bar{b}$ decay.
However, in Refs.~\cite{Htb} it has been demonstrated that the 
 $H^+\to t\bar{b}$ signature can lead to a visible signal at the LHC
provided that the charged Higgs mass is below 600 GeV and 
$\tan\beta$ is either below $\la 1.5$ or above $\ga 40$.
An alternative decay mode to detect a heavy charged Higgs boson is 
$H^\pm\to \tau\nu$  \cite{odagiri}, even if such a decay is 
suppressed for heavy charged Higgs bosons, it has the advantage of being 
much cleaner than  $H^+\to t\bar{b}$.  Recently, a new technique 
using the jet substructure for the heavy charged Higgs boson 
decaying to $tb$ has been proposed in \cite{Yang:2011jk}.

In the MSSM,  the branching ratio of the decay mode
$B(H^\pm \to W^\pm h)$ could at best be at the level of $10$\%
for low $\tan\beta$ while in the 2HDM-I
\footnote{See Section \ref{sec:2hdm} for classification of 2HDMs.}
it could dominate over $B(H^+\to t\bar{b})$. 
Therefore, $H^\pm \to W^\pm h$ could be an
alternative channel to discover the heavy charged Higgs boson at the
LHC \cite{morettineutre}.
Similarly, 
when the CP-odd Higgs boson $A$ is light enough,
the decay of $H^\pm \to W^\pm A$ could be the dominant 
one in the 2HDMs and could also be used to search for heavy charged Higgs
bosons.
Finally, in the models with higher Higgs representations such as 
the triplet representation of the Higgs, 
the charged Higgs boson could decay into 
$W^\pm Z$ with a significant
branching fraction \cite{Chiang:2015kka}. This decay channel could lead to
isolated leptons in the final state and could be used to distinguish between
models with charged Higgs bosons.

The aim of this work is to study singly-charged Higgs boson 
production in association  with a bottom quark and a jet $q'$ with
the subprocess $q b\to q' H^+ b$. 
Such a process had been studied for the first time
in Ref.~\cite{moretti} which showed that the rate is rather small
 in the MSSM due to  a huge cancellation
between the top- and Higgs-mediated diagrams as we will show.
In the present study, we discuss the production rate of this process
and its sensitivity to $\tan\beta$ in 
 the 2HDMs where the masses of the heavier Higgs bosons
are not fixed by one mass parameter as in the MSSM. Specifically,
we demonstrate that the process 
possesses destructive interference between the $s$- and $t$-channel diagrams,
which significantly reduces the cross section. Especially,
when the two heavier neutral Higgs bosons are decoupled
from the lightest one and they are degenerate,
the cross section is canceled to a large extent.
In addition, we show that with a relatively light CP-odd Higgs boson, which is
still allowed by the current data, the production cross section
of the charged Higgs boson 
 via $W^\pm$-Higgs fusion in the $pp \to H^\pm\,b\,j$ process
 can be significantly enhanced at the LHC.

The organization of the work is as follows. 
In the next section, we write down the framework for the 2HDMs,
provide analytic understanding of the process in terms of 
the $2\to 2$ subprocess, 
and also describe the full $2\to 3$ process in detail.
We present the numerical results in Sec.~\ref{sec:num}.
Some cases beyond the 2HDMs
are considered in Sec.~\ref{sec:b2hdm}
and we conclude in Sec.~\ref{sec:con}.

\section{ $q b \to q' H^+ b$ in Two Higgs Doublet Models}
\label{sec:2hdm}
\subsection{Brief review of two-Higgs-doublet models}

In 2HDMs the electroweak symmetry breaking is performed by two 
scalar fields $\Phi_1$ and $\Phi_2$ which are parameterized by
\footnote{ For an overview, see Ref.~\cite{Branco:2011iw}.} :
%
\begin{equation}
\Phi_1=\left(\begin{array}{c}
\phi_1^+ \\ \frac{1}{\sqrt{2}}\,(v_1+\phi_1^0+ia_1)
\end{array}\right)\,; \ \ \
\Phi_2={\rm e}^{i\xi}\,\left(\begin{array}{c}
\phi_2^+ \\ \frac{1}{\sqrt{2}}\,(v_2+\phi_2^0+ia_2)
\end{array}\right) \;.
\end{equation}
We denote $v_1=v \cos\beta=vc_\beta$
and $v_2=v \sin\beta=vs_\beta$. 
The parameterization of the general scalar potential which is gauge 
invariant and possesses a general CP structure can be found in 
\cite{Cheung:2013rva}.
In the present study we are mainly interested in Higgs coupling to fermions
and gauge couplings to be listed slightly later.

The general structure for Yukawa couplings is given 
in the following interactions
\begin{eqnarray}
-{\cal L}_Y&=& h_u\, \overline{u_R}\, Q^T\,(i\tau_2)\,\Phi_2
+h_d\, \overline{d_R}\, Q^T\,(i\tau_2)\,
\left(-\eta_1^d\,\widetilde\Phi_1 -\eta_2^d\,\widetilde\Phi_2\right)
\nonumber \\[2mm] &+&
h_l\, \overline{l_R}\, L^T\,(i\tau_2)\,
\left(-\eta_1^l\,\widetilde\Phi_1 -\eta_2^l\,\widetilde\Phi_2\right)
\ + \ {\rm h.c.}
\end{eqnarray}
where $Q^T=(u_L\,,d_L)$, $L^T=(\nu_L\,,l_L)$, and
$\widetilde\Phi_i=i\tau_2 \Phi_i^*$ with
\begin{equation}
i\tau_2=\left(\begin{array}{cc}
0&1 \\ -1 & 0
\end{array}\right)\,.
\end{equation}
We note that there is a freedom to
redefine the two linear combinations of $\Phi_2$ and $\Phi_1$ to eliminate
the coupling of the up-type quarks to $\Phi_1$~\cite{Davidson:2005cw}. The 2HDMs
are classified according to the values of $\eta_{1,2}^l$ and $\eta_{1,2}^d$
as in Table~\ref{tab:2hdtype}.
%
\begin{table}[!t]
\caption{\label{tab:2hdtype}
Classification of 2HDMs satisfying the Glashow-Weinberg condition
\cite{Glashow:1976nt}
which guarantees the absence of tree-level FCNC.
}
\begin{center}
\begin{tabular}{|l|cccc|}
\hline
\hline
         & \hspace{0.5cm} 2HDM I\hspace{0.5cm} & 2HDM II\hspace{0.5cm}
& 2HDM III\hspace{0.5cm} & 2HDM IV\hspace{0.5cm}  \\
\hline
$\eta_1^d$  & $0$ & $1$ & $0$ & $1$  \\
$\eta_2^d$  & $1$ & $0$ & $1$ & $0$  \\
\hline
\hline
$\eta_1^l$  & $0$ & $1$ & $1$ & $0$  \\
$\eta_2^l$  & $1$ & $0$ & $0$ & $1$  \\
\hline
\hline
\end{tabular}
\end{center}
\end{table}

To define the Higgs mass eigenstates, we first 
rotate the imaginary components $a_i$ and the charged ones $\phi_1^+$ and 
$\phi_2^+$ in order to obtain
the would-be-goldstones $G^0$ and $G^\pm$ that would be eaten
by the longitudinal components of the $Z$ and $W^\pm$ bosons.
These rotations result in an CP-odd state 
 $a=A=-s_\beta a_1+c_\beta a_2$
and a pair of charged Higgs bosons
$H^\pm=-s_\beta \phi_1^\pm +c_\beta \phi_2^\pm$. 
 In the most general case with CP violation,
the mass eigenstates of the neutral
Higgs bosons are obtained by diagonalizing the $3\times 3$ mass matrix 
 ${\cal M}_0^2$
by an orthogonal $3\times 3$ mixing matrix $O$ 
that relates the interaction eigenstates to the mass eigenstates as follow:
\begin{eqnarray}
(\phi_1^0,\phi_2^0,a)^T_\alpha&=&O_{\alpha i} (H_1,H_2,H_3)^T_i
\end{eqnarray}
such that $O^T {\cal M}_0^2 O={\rm diag}(M_{H_1}^2,M_{H_2}^2,M_{H_3}^2)$
with the ordering of $M_{H_1}\leq M_{H_2}\leq M_{H_3}$. 
Here the states $H_i$ do not have to carry any definite
CP-parity and they have both CP-even and CP-odd components.

After identifying the Yukawa couplings by
\begin{equation}
h_u = \frac{\sqrt{2}m_u}{v}\,\frac{1}{s_\beta}\,; \ \
h_d = \frac{\sqrt{2}m_d}{v}\,\frac{1}{\eta_1^dc_\beta+\eta_2^d s_\beta}\,; \ \
h_l = \frac{\sqrt{2}m_l}{v}\,\frac{1}{\eta_1^lc_\beta+\eta_2^l s_\beta}\,,
\end{equation}
one can easily obtain, from the above Lagrangian, the 
following Higgs-fermion-fermion interactions
\begin{eqnarray}
\label{eq:nhff.2hdm}
-{\cal L}_{H_i\bar{f}f} &=&
\frac{m_u}{v}\left[
\bar{u}\, \left(
\frac{O_{\phi_2 i}}{s_\beta} -i\,\frac{c_\beta}{s_\beta}O_{ai}\,\gamma_5\,
\right)\,u \right]\,H_i
\nonumber \\[2mm] &+&
\frac{m_d}{v}\left[
\bar{d}\, \left(
\frac{\eta_1^dO_{\phi_1 i}+\eta_2^dO_{\phi_2
i}}{\eta_1^dc_\beta+\eta_2^ds_\beta}
-i\,\frac{\eta_1^ds_\beta-\eta_2^dc_\beta}{\eta_1^dc_\beta+\eta_2^ds_\beta}
O_{ai}\,\gamma_5\,
\right)\,d \right]\,H_i
\nonumber \\[2mm] &+&
\frac{m_l}{v}\left[
\bar{l}\, \left(
\frac{\eta_1^lO_{\phi_1 i}+\eta_2^lO_{\phi_2
i}}{\eta_1^lc_\beta+\eta_2^ls_\beta}
-i\,\frac{\eta_1^ls_\beta-\eta_2^lc_\beta}{\eta_1^lc_\beta+\eta_2^ls_\beta}
O_{ai}\,\gamma_5\,
\right)\,l \right]\,H_i
\end{eqnarray}
and
\begin{eqnarray}
\label{eq:chff.2hdm}
-{\cal L}_{H^\pm\bar{u}d} &=&
-\frac{\sqrt{2}m_u}{v}
\left(\frac{c_\beta}{s_\beta}\right) \,\bar{u}\,P_L\,d\,H^+
- \frac{\sqrt{2}m_d}{v}\left(\,
\frac{\eta_1^ds_\beta-\eta_2^dc_\beta}{\eta_1^dc_\beta+\eta_2^ds_\beta}
\right)\,\bar{u}\,P_R\,d\,H^+
\nonumber \\ &&
- \frac{\sqrt{2}m_l}{v}\left(\,
\frac{\eta_1^ls_\beta-\eta_2^lc_\beta}{\eta_1^lc_\beta+\eta_2^ls_\beta}
\right)\,\bar{\nu}\,P_R\,l\,H^+
\ + \ {\rm h.c.} \,,
\end{eqnarray}
where $P_{L,R} = (1 \mp \gamma^5)/2$.

Before moving to the next subsection, we present the mixing matrix
$O$ in the CP-conserving case in terms of the mixing angle $\alpha$. 
In our numerical study, to deliver our findings more clearly,
we focus on the CP-conserving case.
In this case
the matrix $O$ takes the following form:
\begin{equation}
\label{eq:o-cpc}
O=\left(\begin{array}{ccc}
-\sin\alpha&\cos\alpha & 0\\
\cos\alpha & \sin\alpha &0\\
0 & 0 & 1
\end{array}\right)\,,
\end{equation}
assuming $H_3$ is the pure CP-odd state or $H_3=A$.
In this notation, the decoupling limit of the 2HDM \cite{Gunion:2002zf},
which seems to be favored by the current LHC data, is $\beta-\alpha\to \pi/2$:

\begin{eqnarray}
&&
O_{\phi_1 1}=-\sin\alpha \to \cos\beta\,, \ \ \
O_{\phi_1 2}= \cos\alpha \to \sin\beta\,; \nonumber \\
&&
O_{\phi_2 1}= \cos\alpha \to \sin\beta\,,  \ \ \
O_{\phi_2 2}= \sin\alpha \to -\cos\beta\,.
\end{eqnarray}

\begin{figure}[t!]
\begin{center}
\includegraphics[width=6.2in]{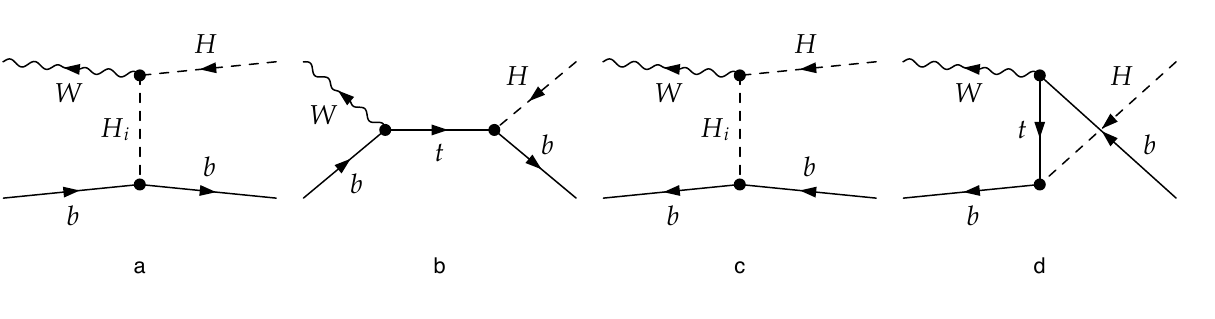}
\vspace{-1cm}
\end{center}
\caption{ \label{fig:subprocess}
 Feynman diagrams for $2\to 2$ subprocesses: 
$W^+b \to  H^+ b$ (a) and (b), 
$W^+\bar{b} \to  H^+ \bar{b}$ (c) and (d).}
\end{figure}

\subsection{Subprocess $W^+\,b\to H^+\, b$ and unitarity}
%
In this subsection,  we present the amplitude of the process 
$q\,b \to q' H^\pm\, b\,  $ in the effective $W$ approximation.
In this process, the dominant contribution comes from the region where
the $W$ boson emitted from the incoming quark $q$ is close to on shell and
one can approximately represent the process by
the $W$ boson scattering with the incoming $b$ quark or anti-$b$ quark
to give $H^\pm b$ or $H^\pm \bar{b}$ in the final state:
\begin{eqnarray}
&&W^+ (q_1) \ b(p_1) \ \to \ H^+ (q_2) \ b (p_2) \,. \nonumber\\
&&W^+ (q_1) \ \bar{b}(p_1) \ \to \ H^+ (q_2) \ \bar{b} (p_2) \,. 
\end{eqnarray}
The process $W^+ b \to H^+ b$ receives contributions from
Fig.~\ref{fig:subprocess}$(a)$ a $t$-channel diagram with 
 the neutral $H_i$ exchanges
 and Fig.~\ref{fig:subprocess}$(b)$ a $s$-channel diagram with top exchange.
 While the process $W^+ \bar{b} \to H^+ \bar{b}$ receives contributions from
Fig.~\ref{fig:subprocess}$(c)$ a $t$-channel diagram with 
 the neutral $H_i$ exchanges
 and Fig.~\ref{fig:subprocess}$(d)$ a $u$-channel diagram with top exchange.
The relevant interactions needed for these two subprocesses can be 
obtained from the Yukawa interactions given by
 Eqs.~(\ref{eq:nhff.2hdm}) and (\ref{eq:chff.2hdm})
and from the covariant derivatives:
\begin{eqnarray}
{\cal L}_{H_i\bar{b}b}&=&-\frac{gm_b}{2m_W}\,
\bar{b}\left(g^S_i+i\,g^P_i\,\gamma_5\right)b\,H_i \,, \nonumber \\[2mm]
{\cal L}_{H^\pm tb}&=&+\frac{gm_b}{\sqrt{2}m_W}\,
\bar{b}\left(c_L\,P_L+c_R\,P_R\right)t\,H^- \ + \ {\rm h.c.}\,, \nonumber 
  \\[2mm]
{\cal L}_{W^\pm tb}&=&-g/\sqrt{2}\,(\bar{t}\gamma_\mu P_L\,b)\,W^{+\mu}+{\rm
h.c.} \,, \nonumber\\
{\cal L}_{H_iH^\pm W^\pm}&=&-\frac{g}{2}\,(S_i + iP_i)\,
\left[H^-\left(i\stackrel{\leftrightarrow}{\partial_\mu}
\right)H_i\right]\,W^{+\mu} \ + \ {\rm h.c.}\,,
\end{eqnarray}
where
\begin{equation}
S_i=c_\beta O_{\phi_2 i} -s_\beta O_{\phi_1 i}\,, \ \ \
P_i=O_{ai}\,,
\end{equation}
and 
\begin{equation}
c_L=\tan\beta\,, \ \ c_R=\frac{m_t}{m_b}\,\frac{1}{\tan\beta}\,; \ \ \
g^S_i=\frac{O_{\phi_1 i}}{c_\beta}\,, \ \ g^P_i=-\tan\beta{O_{ai}}
\end{equation}
in types II and IV and
\begin{equation}
c_L=-\frac{1}{\tan\beta}\,, \ \ c_R=\frac{m_t}{m_b}\,\frac{1}{\tan\beta}\,; \ \ \
g^S_i=\frac{O_{\phi_2 i}}{s_\beta}\,, \ \ g^P_i=\frac{O_{ai}}{\tan\beta}
\end{equation}
in types I and III.

The amplitude of each diagram for 
$ W^+ (q_1)\; b (p_1) \to H^+ (q_2)\; b (p_2) $  reads
\begin{eqnarray}
{\cal M}_{(a)}^{H_i}&=&-\frac{g^2m_b }{4m_W (t-M_{H_i}^2)}\,
(S_i+iP_i)\,(q_2+p_{H_i})^\mu \epsilon_{\mu}(q_1)\,
\left[
\bar{u}(p_2) \left(g^S_i+ig^P_i\gamma_5\right)  u(p_1)
\right]\,, \nonumber \\[3mm]
{\cal M}_{(b)}&=&-\frac{g^2m_b C_v}{2m_W (s-m_t^2)}\,
\left[
c_L\,\bar{u}(p_2)\, \slash{p}_{\!t}\, \slash\epsilon(q_1) P_L u(p_1) \ + \
c_R\,m_t\,
\bar{u}(p_2) \slash\epsilon(q_1) P_L u(p_1) \right]\,,  
\end{eqnarray}
where $s=(p_1+q_1)^2=(p_2+q_2)^2$,
$t=(p_1-p_2)^2=(q_2-q_1)^2$, and
$u=(p_1-q_2)^2=(p_2-q_1)^2$ and $\epsilon^\mu(q_1)$ denotes
the polarization vector of $W^+$ boson.
The amplitudes for the $(c)$ and $(d)$ diagrams in Fig.~\ref{fig:subprocess}
can be obtained by replacing $u(p_{1,2})$ with $v(p_{1,2})$ and
$(s-m_t^2)$ with $(u-m_t^2)$.

In the high-energy limit, $s,|t|,|u| \gg m_W^2, m_t^2, M_{H_i}^2, M_{H^\pm}^2$, 
we find that
\begin{eqnarray}
\label{eq:mab}
&&
{\cal M}_{(a)+(b)}=
\sum_i{\cal M}_{(a)}^{H_i} + {\cal M}_{(b)}
\approx \,
\frac{g^2m_b}{4m_W^2}\,\Bigg\{
\left[
\sum_{i}(S_ig^S_i-P_ig^P_i) \ + \
i\sum_{i}(S_ig^P_i+P_ig^S_i) \,\right] \bar{u}(p_2) P_R u(p_1)
\nonumber \\
&&\hspace{2.0cm}+
\left[
\left(2\,c_L+\sum_{i}(S_ig^S_i+P_ig^P_i)\right) \ + \
i\sum_{i}(-S_ig^P_i+P_ig^S_i) \right]\, \bar{u}(p_2) P_L u(p_1)
\Bigg\} \;,
\end{eqnarray}
where we have taken the longitudinally polarized $W$ or 
$\epsilon^\mu(q_1)\approx q_1^\mu/m_W$:
$q_1^\mu =p_t^\mu-p_1^\mu$ with $p_t^2=s$ for the diagram $(b)$
and
$q_1^\mu =p_t^\mu+p_2^\mu$ with $p_t^2=u$ for the diagram $(d)$,
respectively,
denoting the four--momenta of the exchanging top quark with $p_t$.
Incidentally, the square of the 4-momenta of the internal neutral Higgs is 
$p_{H_i}^2=(p_1-p_2)^2=t$.
We note that the $c_R$ term, which is suppressed by
$m_t/\sqrt{s}$, is neglected here.
In types II and IV, the $c_R$ term 
could be important when $\tan\beta \lsim \sqrt{m_t/m_b}\sim 7$.
As shall be seen,
the total cross section takes its smallest value at $\tan\beta\sim 7$.
When $\tan\beta\gsim 7$, compared to the $c_L$ term,
the $c_R$ term could be safely neglected
when $\sqrt{s}/m_t \gg (m_t/m_b)/\tan^2\beta$.
On the other hand, in types I and III, the $c_R$ term can be neglected
only if $\sqrt{s}/m_t \gg m_t/m_b$. 
Therefore, the high-enegy limit should be applied with more 
cautions at the LHC for types I and III.  But, for the 2HDM types I and III,
the production cross sections are
suppressed by $1/\tan^2\beta$ with increasing $\tan\beta$ and the largest value 
with $\tan\beta=1$ is only $\sim 30$ fb, as shall be shown.

The amplitude ${\cal M}_{(a)+(b)}$ for the $b$-initiated processes
consists of the contributions from
the $t$-channel Higgs-exchange diagrams $(a)$ and
the $s$-channel top-exchange diagram $(b)$ .
The $c_L$ term in the second line is from the $s$-channel
diagram and all the others from the $t$-channel ones.
Therefore, the high-energy limit has been obtained by taking
$s/(s-m_t^2)\approx t/(t-M_{H_i}^2) \approx 1$.
On the other hand, the high-energy limit of the 
amplitude ${\cal M}_{(c)+(d)}$ for the $\bar{b}$-initiated processes
can be obtained by replacing $u(p_{1,2})$ with $v(p_{1,2})$
in Eq.~(\ref{eq:mab}) and 
taking $u/(u-m_t^2)\approx t/(t-M_{H_i}^2) \approx 1$.
We note that the high-energy behavior of ${\cal M}_{(a)+(b)}$ 
is different from that of ${\cal M}_{(c)+(d)}$
especially when $s$ is not large enough and
there are non-negligible parts of phase space in which 
the high-energy limits $u/(u-m_t^2) \approx t/(t-M_{H_i}^2) \approx 1$
in ${\cal M}_{(c)+(d)}$ are much more difficult to achieve 
than the corresponding ones 
$s/(s-m_t^2) \approx t/(t-M_{H_i}^2) \approx 1$
in ${\cal M}_{(a)+(b)}$.
Otherwise, the expression given by Eq.~(\ref{eq:mab}) can be applicable
for both the $b$- and $\bar{b}$-initiated processes.

%
%

The high-energy limit expression
Eq.~(\ref{eq:mab}) contains two non-interfering terms
both of which grow as $\sqrt{-t}$
and therefore the absence of these
unitarity-breaking terms require the following three types of sum rules:
\begin{eqnarray}
\label{eq:sumrules}
&&
2\,c_L+\sum_{i}(S_ig^S_i+P_ig^P_i)=0\,,\nonumber \\
&&
\sum_{i}S_ig^S_i=\sum_{i}P_ig^P_i\,,\nonumber \\
&&
\sum_{i}S_ig^P_i=\sum_{i}P_ig^S_i=0\,.
\end{eqnarray}
The first one gives the relation between the charged Higgs
coupling to $t$ and $b$ quarks ($c_L$) and the sum over the Higgs states
of the scalar and pseudoscalar products ($g_i^S S_i+g_i^P P_i$)
of the neutral Higgs couplings to $b$ quarks 
and those to the charged Higgs and $W$.
The second relation shows the sum over the Higgs states of
the scalar products should be the same as that of 
the pseudoscalar ones.
And the third relation implies that there is no CP violation if the
scalar-pseudoscalar products are summed over the three Higgs states.

These interesting sum rules can be explicitly checked in each
2HDM. In types II and IV,  using the orthogonality of the mixing matrix $O$,
we find that
\begin{eqnarray}
\sum_{i}S_ig^S_i&=&\sum_{i}(O_{\phi_2 i}O_{\phi_1 i}-\tan\beta O_{\phi_1 i}^2)
=-\tan\beta\,, \nonumber \\[2mm]
\sum_{i}P_ig^P_i&=&-\tan\beta \sum_{i}O_{a i}^2 =-\tan\beta\,,  \nonumber \\[2mm]
\sum_{i}S_ig^P_i&=&\sum_{i}P_ig^S_i=0 \;.
\end{eqnarray}
With $c_L=\tan\beta$, the unitarity conditions are satisfied automatically.
On the other hand, in types I and III, we find that
\begin{eqnarray}
\sum_{i}S_ig^S_i&=&\sum_{i}(\frac{O_{\phi_2 i}^2}{\tan\beta}
-O_{\phi_1 i}O_{\phi_2 i}) =1/\tan\beta\,, \nonumber \\[2mm]
\sum_{i}P_ig^P_i&=&\sum_{i}\frac{O_{a i}^2}{\tan\beta} =1/\tan\beta\,, 
\nonumber \\[2mm]
\sum_{i}S_ig^P_i&=&\sum_{i}P_ig^S_i=0\,.
\end{eqnarray}
With $c_L=-1/\tan\beta$, the unitarity conditions are again satisfied
automatically.

This is the proof for the unitarity of the subprocess 
$W^+b\to bH^+$ in the high energy limit in the general 2HDMs 
with or without CP violation. 
The same proof also applies to the case of $\bar{b}$ initiated subprocess
$W^+\bar{b}\to \bar{b}H^+$.

\subsection{The full process $q b \to  q' H^+ b$ }

\begin{figure}[t!]
\begin{center}
\includegraphics[width=6.2in]{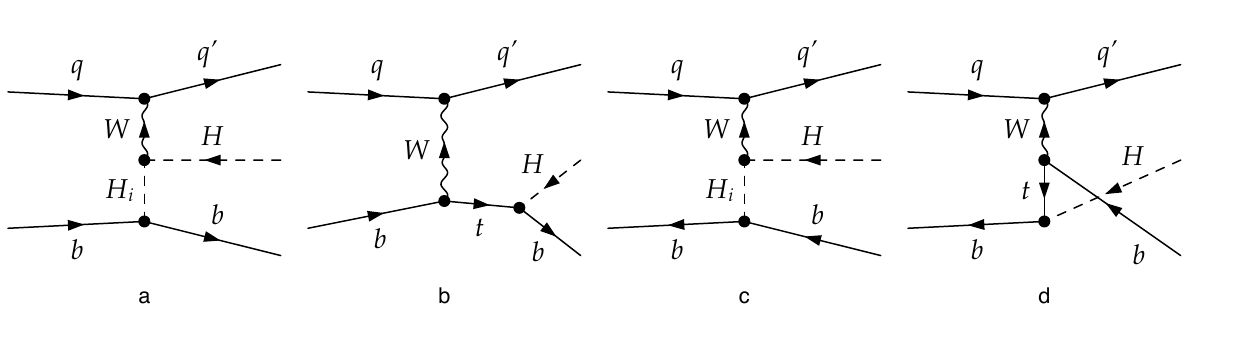}
\vspace{-1cm}
\end{center}
\caption{ \small 
\label{fig:process} 
Feynman diagrams for $q b \to q' H^+ b$ (a) and (b), 
$q \bar{b} \to q' H^+ \bar{b}$ (c) and (d) 
where $(q,q')=(u,d), (c,s)$
and $H_i=h, H, A$.
The processes with $(\bar{q},\bar{q}')=(\bar d,\bar u), (\bar s,\bar c)$
are understood.
}
\end{figure}

After discussing the essence of the physics involved in the $2\to2$ 
subprocess, we shall describe the full $2\to 3$ process
\footnote{
For a full consideration of NLO corrections, one may need to take account of
the $2\to 4$ process: $qg \to q H^+ b \bar b$. 
We leave this part for further work.}. 
We shall consider the CP-conserving case for simplicity,
unless stated otherwise.
In this case, without loss of generality, 
we identify $H_1=h$, $H_2=H$, and $H_3=A$,
where $h$ and $H$ denote the lighter and heavier CP-even Higgs bosons,
respectively, and $A$ the CP-odd one. 
The Feynman diagrams for the subprocesses $q b \to q' H^+ b$ and 
$q \bar{b} \to q' H^+ \bar{b}$ are shown in
Fig.~\ref{fig:process}.  We stress at this level one
important difference between the bottom-initiated diagram
in Fig.~\ref{fig:process}(b) and anti-bottom-initiated one in
Fig.~\ref{fig:process}(d) is that 
the former has a $s$-channel exchange top propagator
while the latter has a $u$-channel one.
Similarly, the fermion-line direction of the $q$ can be reversed to include
$\bar q \to \bar q'$ transition. Therefore, we have a number of initial
states for production of $H^+$: 
$ (u, c, \bar d, \bar s )\otimes (b, \bar b)$.
We can then take the charge conjugate to obtain the $H^-$ processes. 

The diagram in Fig.~\ref{fig:process}(b) represents a 
top-induced process. If $M_{H^\pm} < m_t - m_b$, the top quark is 
produced on-shell, then followed
by its decay into $b H^+$. This diagram is entirely dominant over the 
other diagrams.  However, when  $M_{H^\pm} > m_t - m_b$ the top quark is
off-shell, and thus other diagrams also make significant contributions.
In other diagrams, the charged Higgs boson appears being produced
by $W H_i$ fusion, where $H_i=h, H, A$ in the CP-conserving case. 
The coupling in the
vertex $W^+ H^- H_i$ is a gauge coupling proportional to $g$ and some mixing
angles of the Higgs sector, and
independent of different types of 2HDMs. 
On the other hand, the dependence on the type of 2HDMs comes from the
Yukawa couplings of $H_i$ to $b$ quark and the charged Higgs boson 
to $tb$. We list the relevant Yukawa couplings for 2HDMs from type I to IV
in Table~\ref{tab:1to4} up to some normalizations. 
Incidentally, the non-vanishing
neutral Higgs couplings to charged Higgs and $W$ are given by
\begin{eqnarray}
S_1 &=& S_h=c_\beta O_{\phi_2 1} -s_\beta O_{\phi_1 1}=\cos(\beta-\alpha)
\,, \nonumber \\
S_2 &=& S_H=c_\beta O_{\phi_2 2} -s_\beta O_{\phi_1 2}=-\sin(\beta-\alpha)
\,, \nonumber \\
P_3 &=& P_A=O_{a3}=1\,,
\end{eqnarray}
using the form of $O$ given by Eq.~(\ref{eq:o-cpc}).
%
\begin{table}[t!]
\caption{\small \label{tab:1to4}
The bottom quark Yukawa couplings for $h, H, A$ and that of the charged
Higgs boson for 2HDMs of type I, II, III, and IV. The common factor 
for neutral Higgs boson is $g m_b / \sqrt{2}M_W$ while that for the 
charged Higgs is $g /\sqrt{2} M_W$. The chiral projection operators
are $P_{L,R} = (1 \mp \gamma^5)/2$.
}
\begin{ruledtabular}
\begin{tabular}{lcc}
  & Type I, III  & Type II, IV \\
\hline
$hb\bar b$  & $\frac{\cos\alpha}{\sin\beta}$ & $-\frac{\sin\alpha}{\cos\beta}$
  \\
$H b\bar b$ & $\frac{\sin\alpha}{\sin\beta}$ & $\frac{\cos\alpha}{\cos\beta}$
    \\
$A b\bar b$  &  $+ \cot\beta$ & $-\tan\beta$ \\
 $H^- t \bar{b}$
&  $- \frac{m_b}{\tan\beta} P_L + \frac{m_t}{\tan\beta} P_R$ &
   $ m_b\tan\beta P_L + \frac{m_t}{\tan\beta} P_R$ \\
\end{tabular}
\end{ruledtabular}
\end{table} 

In the decoupling limit, we have
$\cos(\beta - \alpha)=0$ and $\sin(\beta - \alpha) =1$.  
The contribution
from the light Higgs $h$ diagram is automatically zero because $S_h=0$.
The contributions from $H$ and $A$ are the same 
up to the $\gamma^5$ factor in the $\phi^0 b \bar b$ vertex 
if they are degenerate.
If we look at the diagram more closely, the whole process can be
regarded as $W b$ and $W\bar{b}$ annihilation, as $2\to 2$ processes.
It is easy to see from Fig.~\ref{fig:process} that for the $W b\to H^+b$ 
subprocess we have three $t$-channel diagrams with $H_i=h,H,A$ 
in Fig.~\ref{fig:process}(a) and one $s$-channel diagram 
mediated by the top quark in Fig.~\ref{fig:process}(b). 
Similarly, for $W \bar{b}\to H^+\bar{b}$ 
subprocess we have three $t$-channel diagrams with $H_i=h,H,A$ 
in Fig.~\ref{fig:process}(c) and one $u$-channel diagram 
mediated by the top quark in Fig.~\ref{fig:process}(d).
We have shown in the previous subsection
using the effective $W$ approximation that there is strong cancellation
among the diagrams, and indeed 
all four diagrams
will exactly cancel one another in the high energy limit.
Therefore, if we employ a much lighter CP-odd Higgs boson, which is still
allowed by the current data, we expect a strong enhancement to the 
production cross section of this process.  Experimentally, one can use
this process to search for the charged Higgs boson and investigate the
effects of light CP-odd Higgs boson.  Perhaps, a negative search would
close out the entire window of light CP-odd Higgs boson.

\section{Numerical results}
\label{sec:num}
In this section, we first present some numerical results for the 
subprocesses $W^+b\to bH^+$  and  $W^+\bar{b}\to \bar{b}H^+$
for a given value of center-of-mass energy $\sqrt{S}$ and then 
consider the full process $pp\to H^+b j$ in the 2HDM of 
type I (III) and II (IV).

\subsection{The $2\to 2$ subprocess in the effective $W$ approximation}
We shall limit ourself to the CP conserving case 
 taking $H_1=h$, $H_2=H$, and $H_3=A$.
And the couplings of the neutral Higgs bosons to 
the charged Higgs and $W$ are:
$S_h=\cos(\beta-\alpha)$,
$S_H=-\sin(\beta-\alpha)$, and 
$P_A=1$. 
Neglecting the contribution from the
lightest Higgs boson $h$ as in the decoupling limit
$\cos(\beta-\alpha)\to 0$, we
observe that in the high-energy limit the cross section of the subprocess
behaves like
\begin{equation}
\label{22}
\sigma(W^+b\to H^+b) \propto
\left|2\,c_L+ S_H g^S_H+ P_A g^P_A\right|^2 + 
\left| S_H g^S_H- P_A g^P_A\right|^2 \;.
\end{equation}
We note that the cross section suffers a huge cancellation
between the top- and Higgs-mediated diagrams
and a further cancellation between the
Higgs-mediated diagrams.
Taking the type II model as an example, we find
\begin{eqnarray}
&&
\left.\sigma(W^+b\to H^+b)\right|_{t\,\,{\rm only}} \propto 
4\,\tan^2\beta\,, \nonumber \\[2mm]
&&
\left.\sigma(W^+b\to H^+b)\right|_{t+H\,\,{\rm only}} =
\left.\sigma(W^+b\to H^+b)\right|_{t+A\,\,{\rm only}} \propto 
2\,\tan^2\beta\,, \nonumber \\[2mm]
&&
\left.\sigma(W^+b\to H^+b)\right|_{t+H+A} \propto 
{\cal O}\left(\frac{m_t^2}{s}+\frac{M_{H_i}^2}{t}\right) 
\label{eq:cancel}
\end{eqnarray}
with $c_L=-S_Hg^S_H=-P_Ag^P_A=\tan\beta$. 
Note, for the $W^+\bar b \to H^+\bar  b$ process,
\begin{equation}
\left.\sigma(W^+\bar b\to H^+\bar b)\right|_{t+H+A} \propto
{\cal O}\left(\frac{m_t^2}{u}+\frac{M_{H_i}^2}{t}\right)
\end{equation}
while 
the high-energy behavior of 
the $t$-only, $(t+H)$-only, and $(t+A)$-only
amplitudes remains the same.

\begin{figure}[t!]
\includegraphics[width=3.2in]{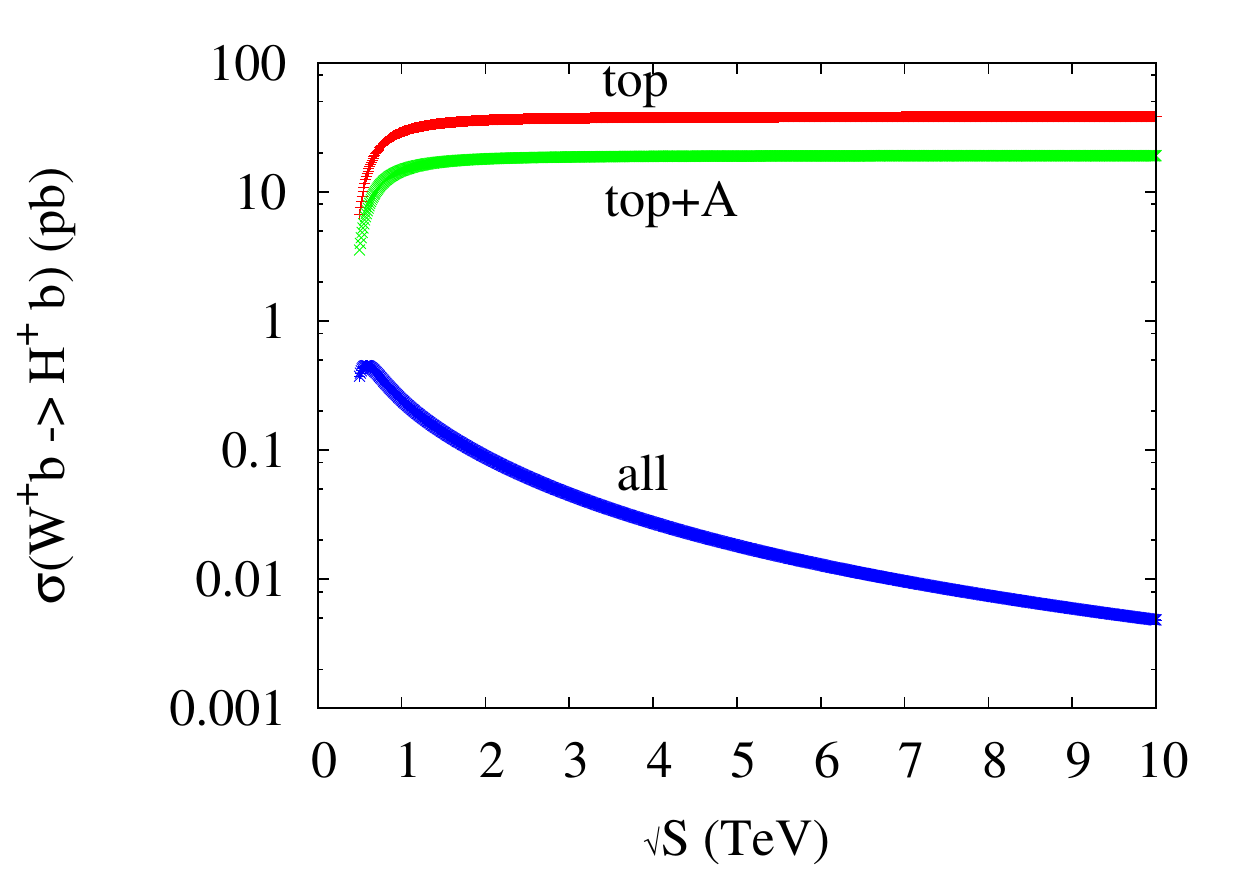}
\includegraphics[width=3.2in]{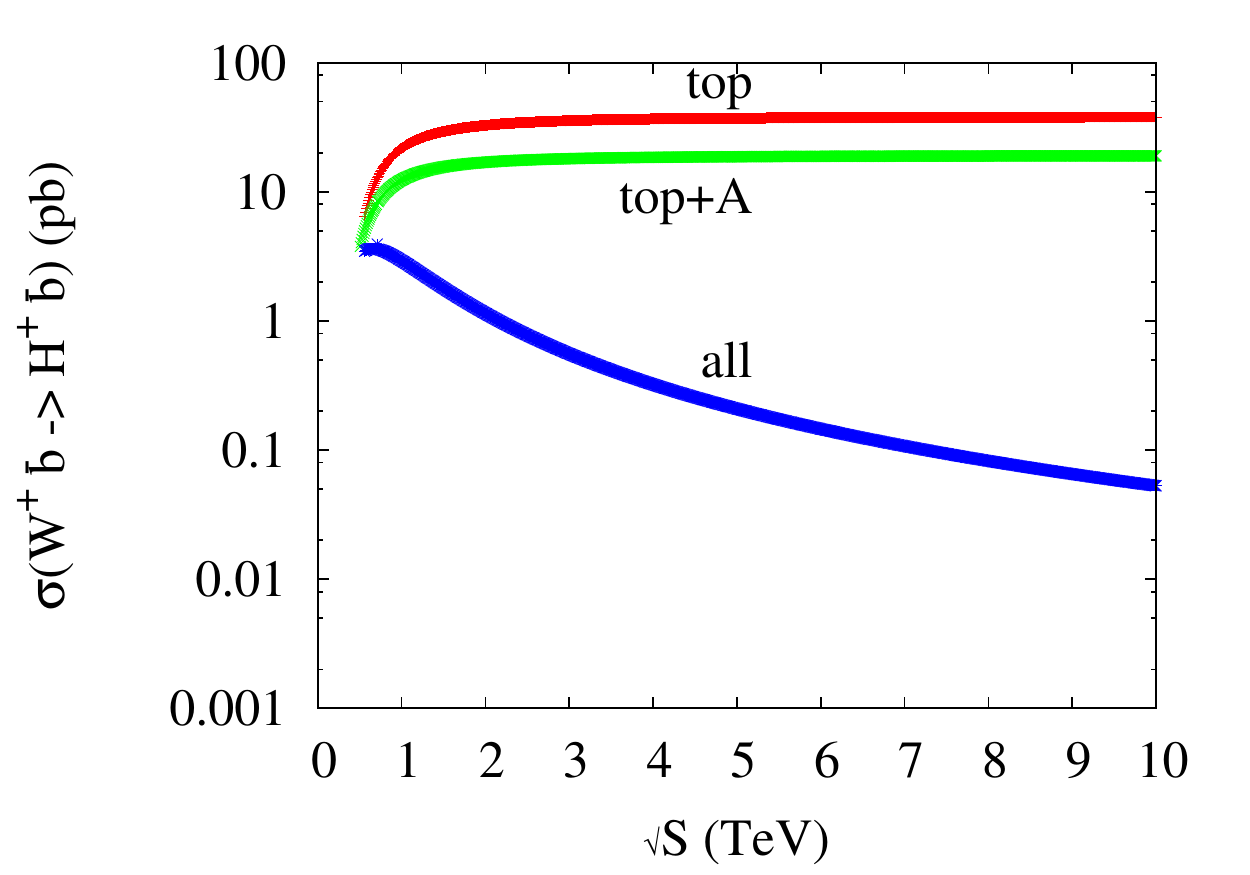}
\caption{\small \label{fig:wbtobhp}
The cross section as a function of center of mass energy for the 
subprocess $W^+ b \to H^+ b$ (left) and $W^+ \bar{b} \to H^+ \bar{b}$ (right)
 in the MSSM for $\tan\beta=30$ and $M_A=400$ GeV.
}
\end{figure}
Furthermore, independent of the type of 2HDMs
we note that for $W^+ b\to H^+ b$  (respectively $W^+ \bar{b}\to H^+ \bar{b}$)
the $s$-channel (respectively the $u$-channel) top-exchange diagram
interferes destructively with the $t$-channel Higgs-exchange 
$W-A$ and $W-H$ fusion diagrams.
For demonstration we show in Fig.~\ref{fig:wbtobhp} the cross sections
for the subprocess $W^+ b\to H^+ b$ (left) and 
$W^+ \bar{b}\to H^+ \bar{b}$ (right) 
as a function of center of mass energy 
 $\sqrt{s}$ in the MSSM. 
\footnote
{
Though we are working in the framework of 2HDMs, 
$M_{H^+}$, $M_H$, and $M_A$ are very close to one another in the MSSM
that will suit our purpose here.
}
We illustrate separately the 
top diagram alone, the sum of the top and pseudoscalar Higgs exchange 
diagrams, as well as all four diagrams.
Note that ``top+$A$'' and ``top+$H$'' are extremely close to each other.
It is clear from the plot that the
dominant contribution is coming from the top diagram. 
It is also visible from the plot that 
the interference between $s$-channel top diagram and $W$-$A$ fusion
 is destructive. The top contribution is reduced by a factor of 2 by
 the $W$-$A$ fusion diagram and same destructive interference takes place with 
 $W$-$H$ fusion diagram. 
We only show the sum of the top diagram and $W$-$A$ fusion diagram
in the figure,
that of the top and $W$-$H$ fusion diagrams is almost the same.
As expected from Eq.~(\ref{eq:cancel}), in the case of $W^+ b\to H^+ b$,
after inclusion of all diagrams the total cross section
drops by more than 3 orders of magnitude at large $\sqrt{s}$,
as shown on the left panel of Fig.~\ref{fig:wbtobhp}.
This in fact is due to the 
strong destructive interference of top diagram with the $W$-$A$ and $W$-$H$ 
fusion diagrams.
Similarly, on the right panel in Fig.~\ref{fig:wbtobhp}
 we illustrate the cross section
for $W^+\bar{b}\to H^+ \bar{b}$ as a function of 
$\sqrt{s}$. 
Again, as
expected we can see destructive interference between $u$-channel 
top diagram and $t$-channel $W$-$A$ and $W$-$H$ fusion diagrams. 
We stress that 
the destructive interference in the $\bar{b}$-initiated process 
is less severe than the $b$-initiated one, 
such that the total cross section 
for $W^+\bar{b}\to H^+ \bar{b}$  is about one order of magnitude larger than
that for $W^+ b\to H^+ b$.
This is because 
the cancellation between the $u$- and $t$-channel diagrams is not as
effective as in the $s$- and $t$-channel diagrams.
For fixed and relatively small values of $\sqrt{s}$,
there are non-negligible parts of phase space in which the high-energy limits
$u/(u-m_t^2) \approx 1$ and $t/(t-M_{H_i}^2) \approx 1$
can not be achieved  simultaneously
due to the relation $t+u = -s+M_W^2 + M_{H^\pm}^2$.
Therefore, the $\bar b$-initiated process has an order of
magnitude larger cross section.

\subsection{For full process $pp\to H^\pm b j$ }
\begin{figure}[t!]
\includegraphics[width=3.2in]{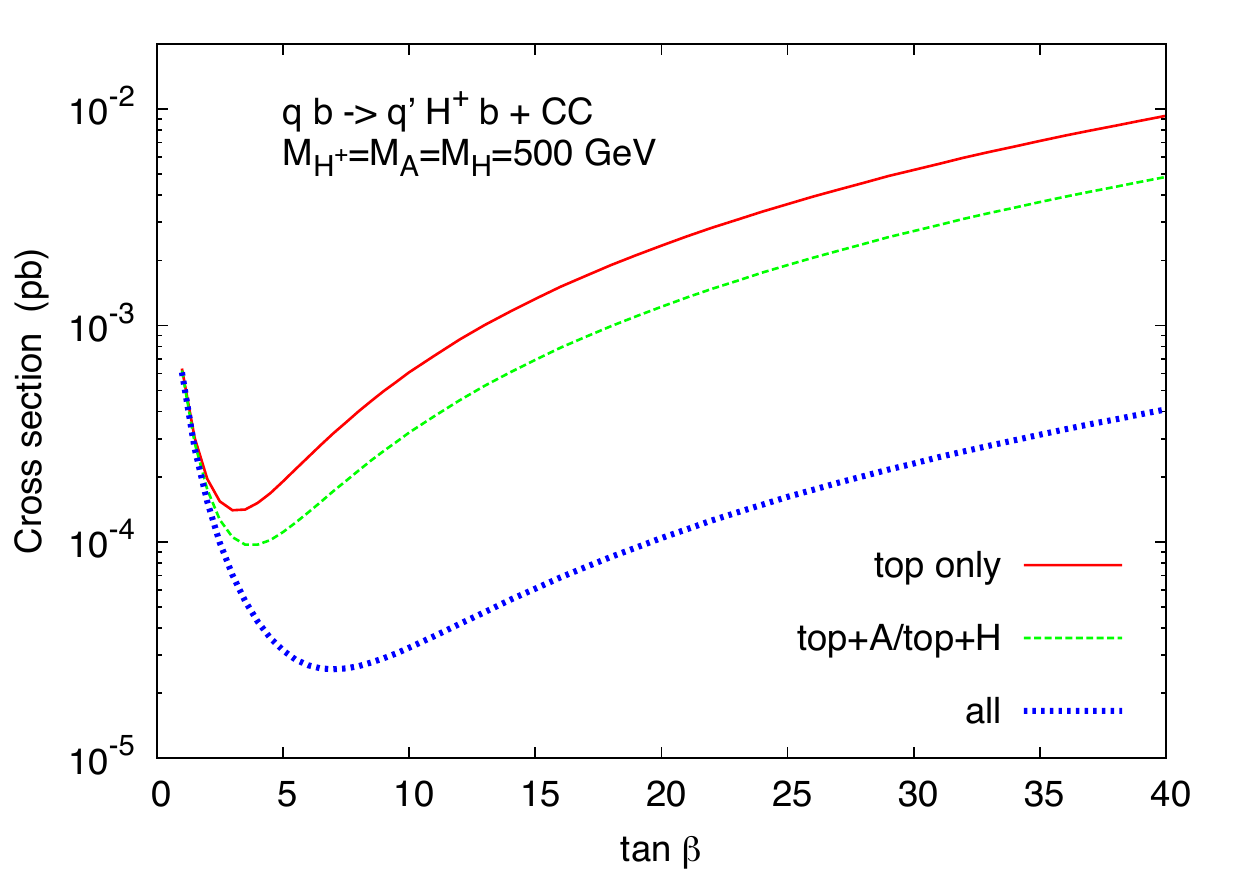}
\includegraphics[width=3.2in]{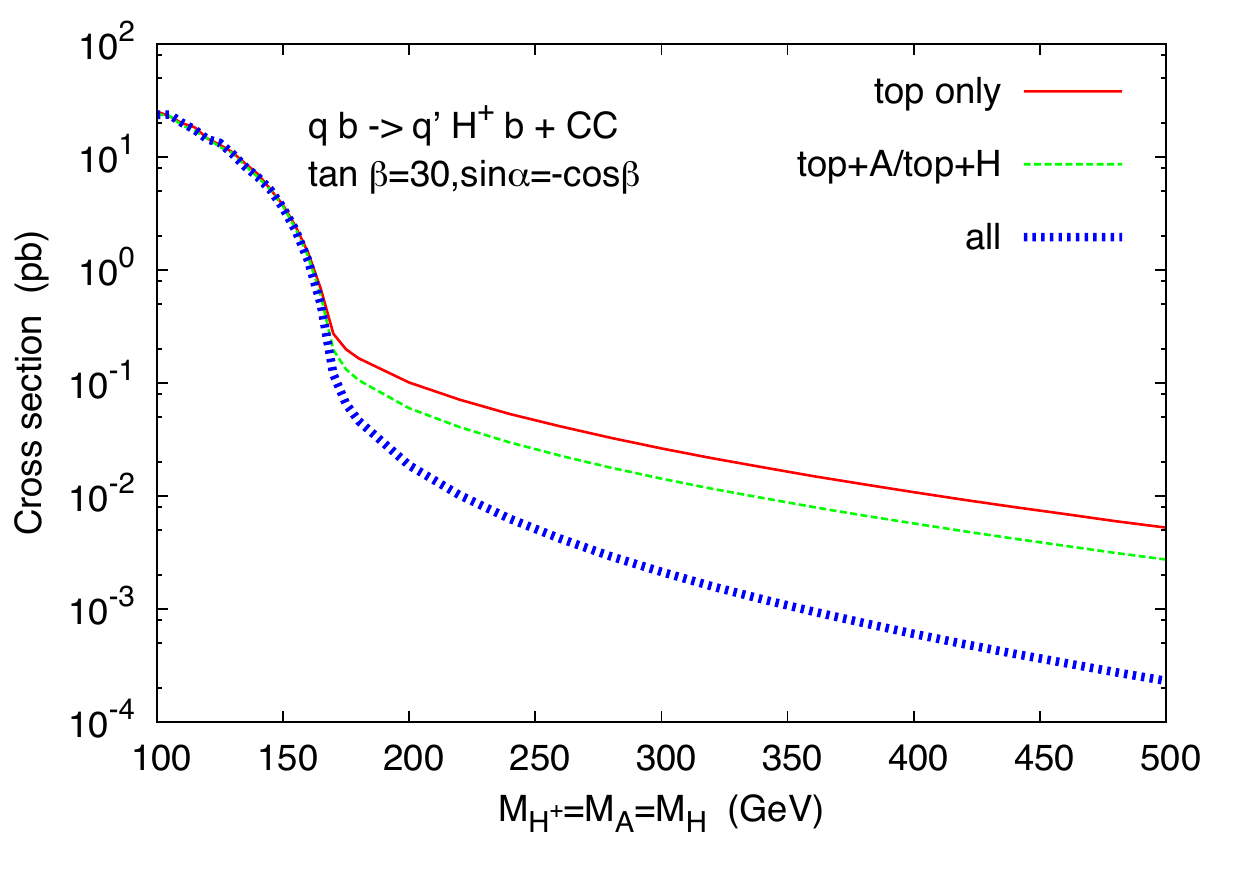}
\includegraphics[width=3.2in]{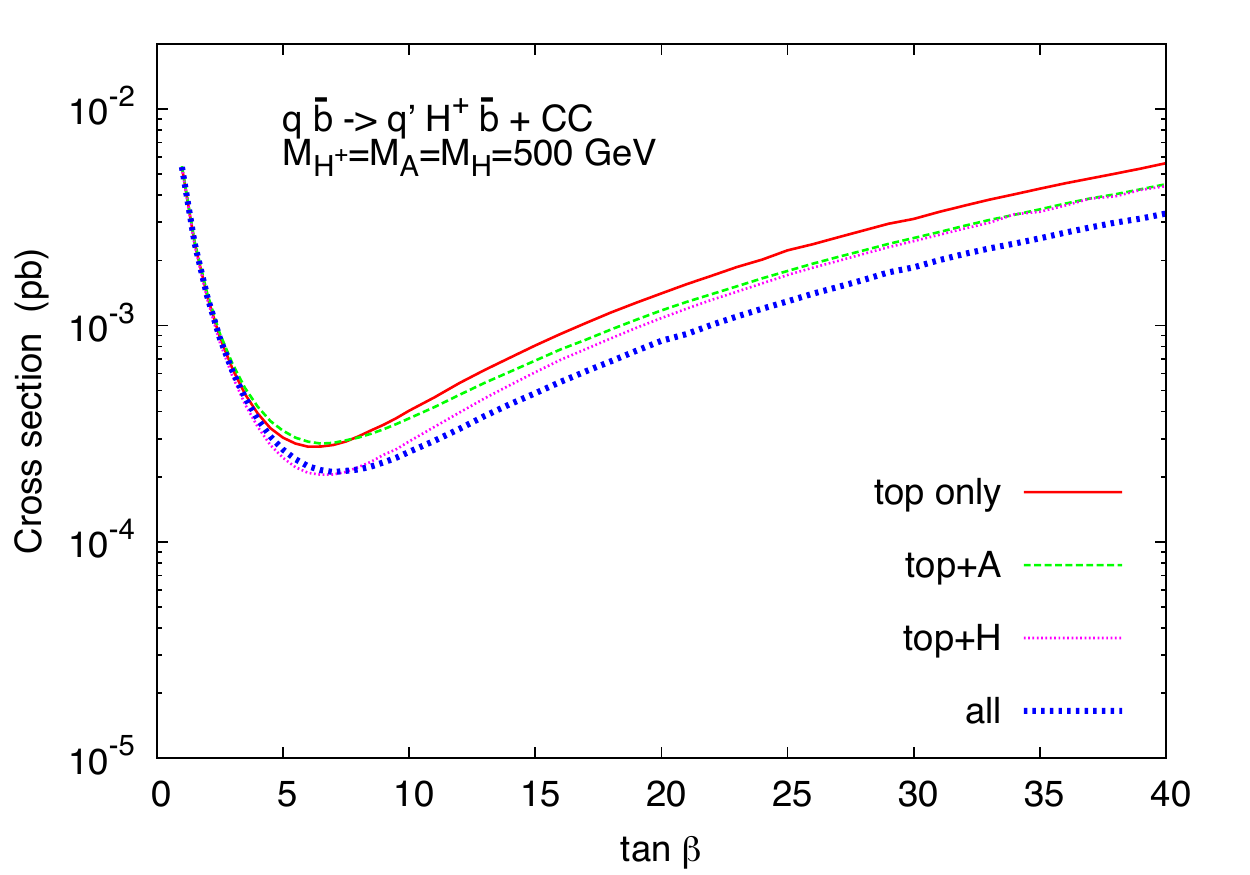}
\includegraphics[width=3.2in]{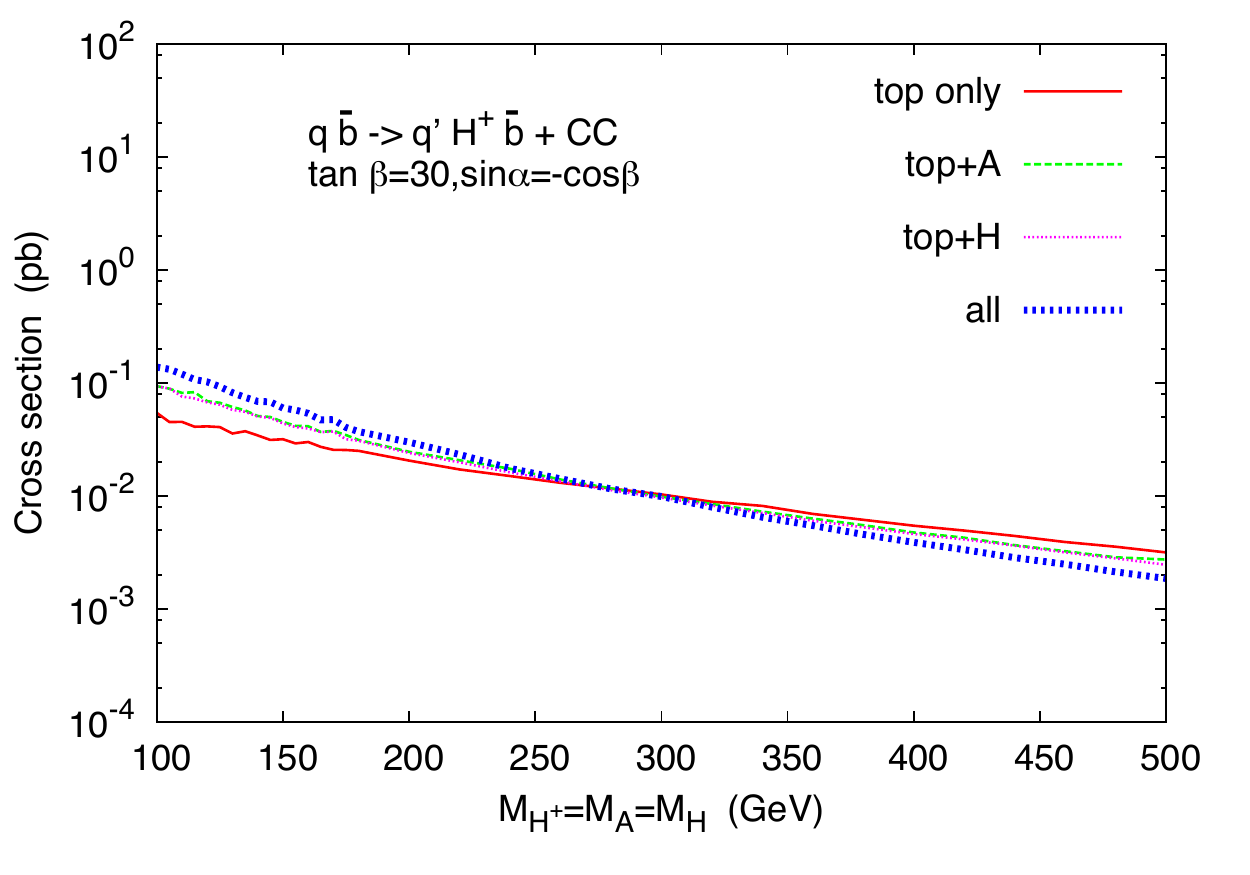}
\includegraphics[width=3.2in]{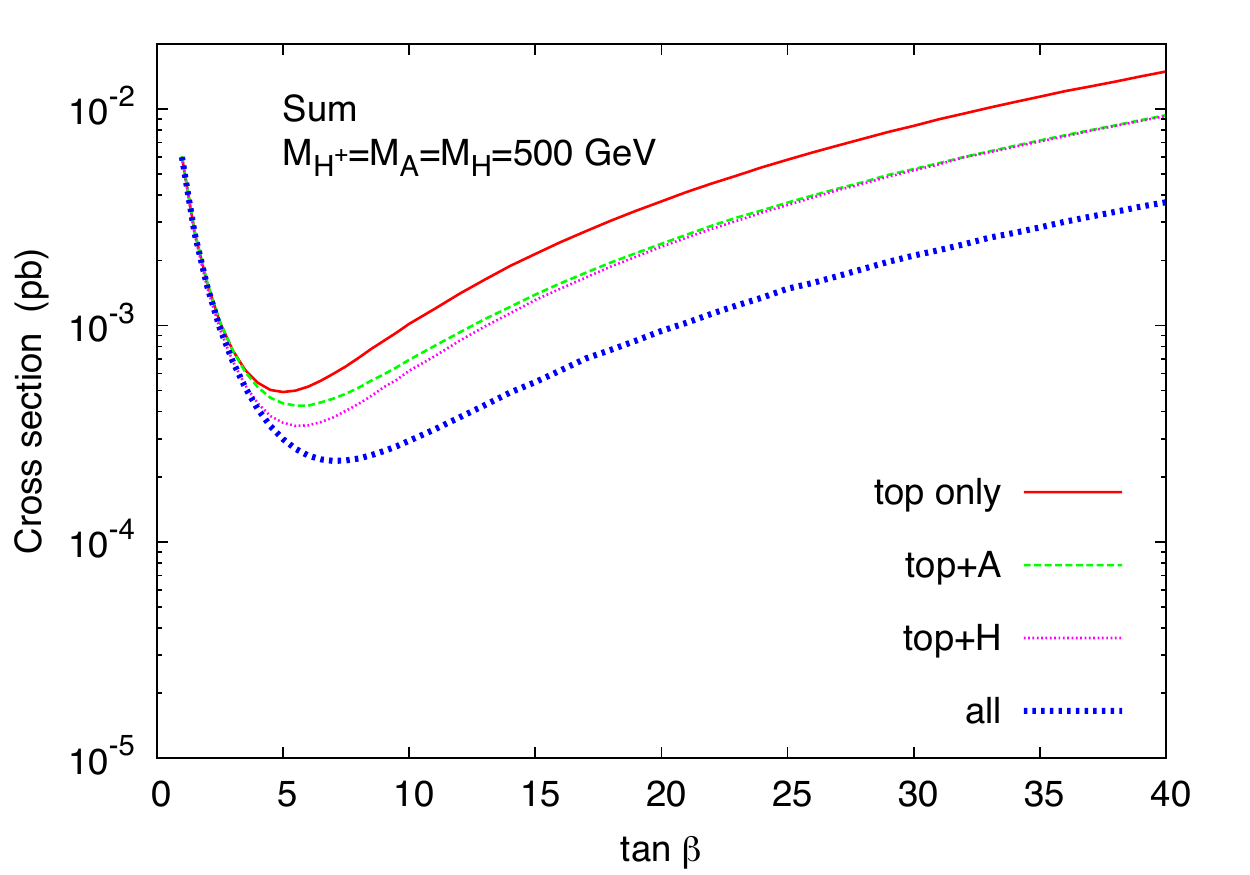}
\includegraphics[width=3.2in]{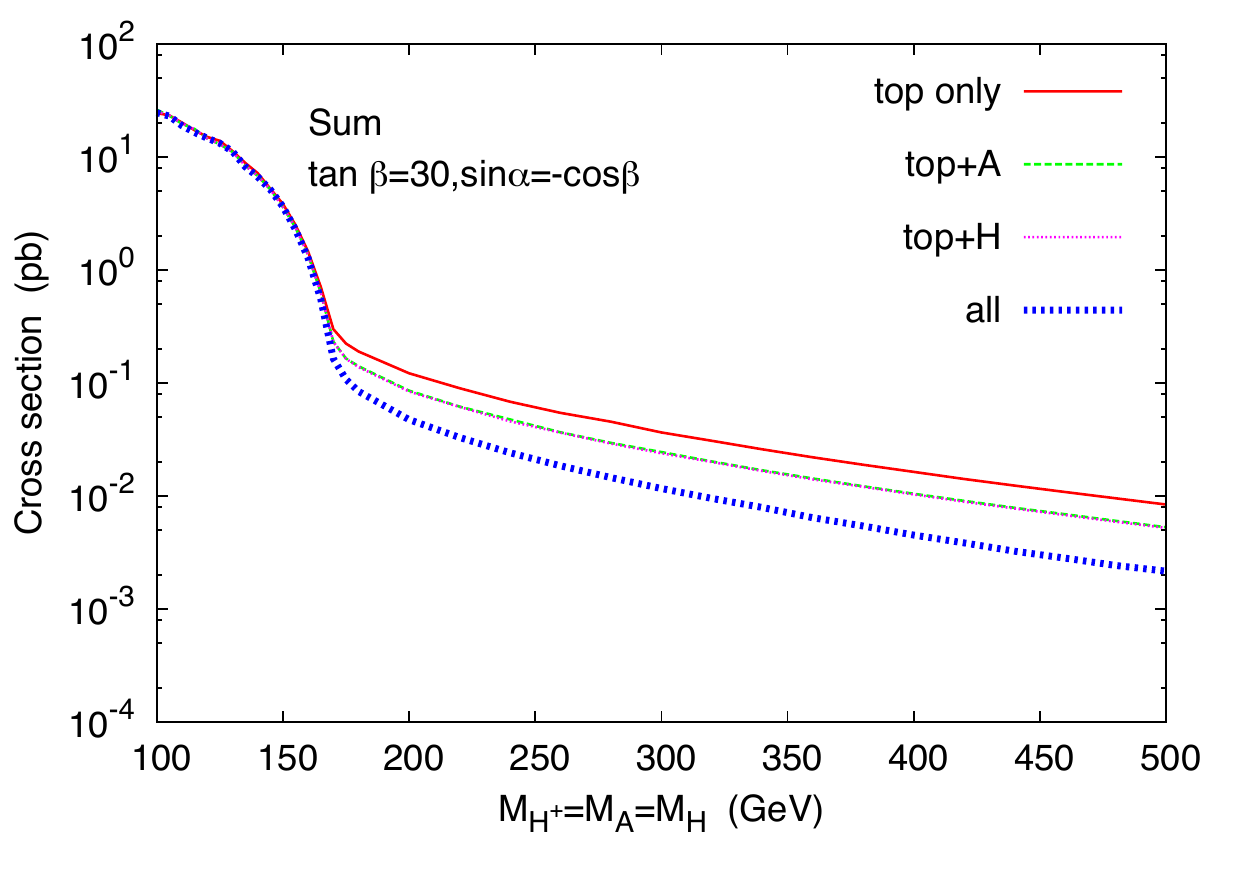}
\caption{\small \label{fig:cx}
The $pp\to H^\pm b j$ cross sections in the 2HDM type II as functions of
$\tan\beta$ (left panels) and charged Higgs mass (right panels) at LHC-14. 
The upper panels are the $b$ initiated process, 
the middle panels are $\bar{b}$ initiated process
while the lower panel are the sum of $b$ and $\bar{b}$.
The charged conjugate panels are included in the plots.
All panels are for the decoupling limit $\sin\alpha=-\cos\beta$. }
\end{figure}

In the previous subsection we have shown analytically 
and illustrated numerically the
cancellation in the subprocesses $W^+b\to b H^+$ and 
$W^+\bar{b}\to H^+\bar{b}$ between the top diagram and 
$W$-$A$ and $W$-$H$ fusion diagrams using the
effective W approximation. 
In Fig.~\ref{fig:cx}, 
we show the cross sections for the full $2\to 3$ processes 
$q b \to q' H^+ b$ (upper panels),
$q\bar b \to q' H^+ \bar b$ (middle panels), and their sum (lower panels)
as functions of $\tan\beta$ (left panels) and $M_{H^+}=M_A=M_{H}$ (right
panels),
including the charged-conjugate channels and 
after folding with the parton distribution functions
\footnote{
Our numerical calculations of the several cross sections for the 
$b$- and $\bar{b}$-initiated full $2\to 3$ processes presented here
are carried out by use of the 
Helicity Amplitude Method~\cite{Barger:1991vn}.
We compare our results for the total cross sections
with those obtained using {\tt MadGraph}~\cite{Alwall:2014hca}
and find excellent agreements.}
Again, we separately show the contributions from the top diagram only,
the top plus $W$-$A$ fusion diagrams, 
the top plus $W$-$H$ fusion diagrams, 
and all diagrams. 
We have assumed that we are in the decoupling limit 
$\sin\alpha=-\cos\beta$ and taking a spectrum of degenerate Higgs bosons
$M_{H^\pm}=M_H=M_A$ as in the MSSM and 
the lightest CP-even Higgs boson $h$ is the observed one 
with $m_h=125.09$ GeV.
Thus, the diagram with $h$ proportional to $\cos(\beta-\alpha)$
does not contribute while the  amplitudes associated with the 
$A$ and $H$ diagrams are the same up 
to a factor of $\gamma^5$ in the $Ab\bar b$ and $Hb\bar b$ vertex.
Also, we can see that in the $b$-initiated subprocess (upper panels) 
the ``top$+A$'' curve completely
overlaps with ``top$+H$'' curves but not exactly in the $\bar b$-initiated
one (middle panels).

The upper panels in Fig.~\ref{fig:cx}
illustrate a very strong cancellation between the top diagram, and 
$W$-$A$ and $W$-$H$ fusion diagrams.  
Note the charged-conjugate channels
$\bar q \bar b \to \bar q' H^- \bar b$ are included in it.
On the other hand, the middle panels show a less severe 
cancellation between the top diagram, 
and $W$-$A$ and $W$-$H$ fusion diagrams where 
the charged-conjugate channels $\bar q b \to \bar q' H^- b$ are also 
included in it.
Therefore, we still see that a strong cancellation occurs
for the full process $pp\to H^\pm b j$ at the LHC-14, shown in the
lower panels.
Also, note that the total cross section is dominated by the $\bar b$-initiated
process when $M_{H^+} \agt  200$ GeV, where the internal top cannot be produced
on-shell.

It is clear from the left panels that the
cross sections are enhanced for both small $\tan\beta\approx 1$ and 
large $\tan\beta$, the latter of which is associated with 
enhanced bottom Yukawa couplings.
A dip indeed occurs around $\tan\beta\approx 
6$ in the $\tan\beta$ plots, 
which corresponds to where the top and bottom Yukawa couplings become similar
$m_b\tan\beta \approx m_t/\tan\beta$.
We stress that our results are in good agreement with Ref.~\cite{moretti}.

As shown on the right panels, we emphasize that for $M_{H^\pm}\leq m_t-m_b$
the top quark can be produced on-shell as in single-top production
and then decays into $b H^+$, Therefore, in the 
range of $M_{H\pm}\leq m_t-m_b$ the top-exchange diagram 
completely dominates over other diagrams.
On the other hand, 
for  $M_{H^\pm} > m_t - m_b$ the top quark is
off-shell, and thus other diagrams also make significant contributions.
Note that in the $\bar b$-initiated process $q \bar b \to q' H^+ \bar b$
the top quark is never produced on-shell.

\begin{figure}[t!]
\includegraphics[width=3.2in]{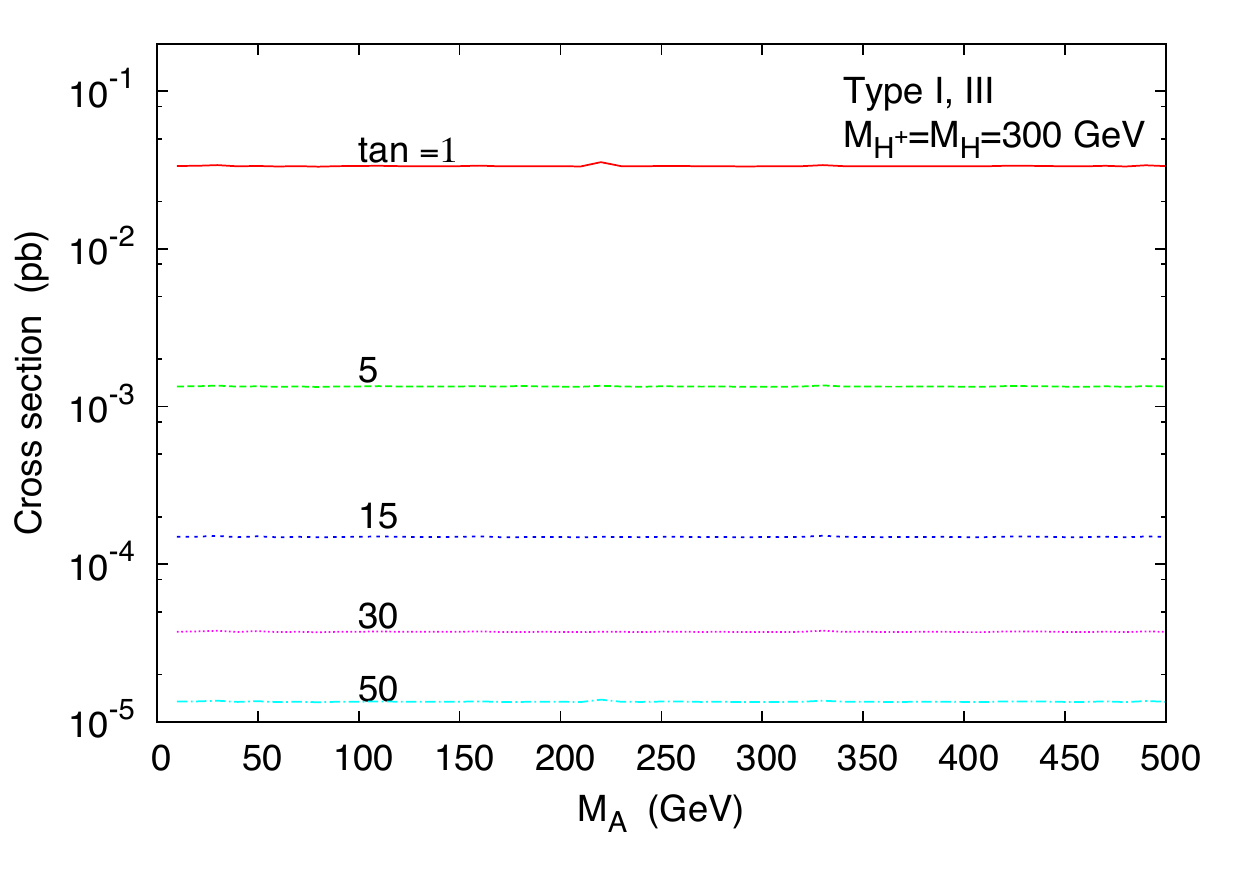}
\includegraphics[width=3.2in]{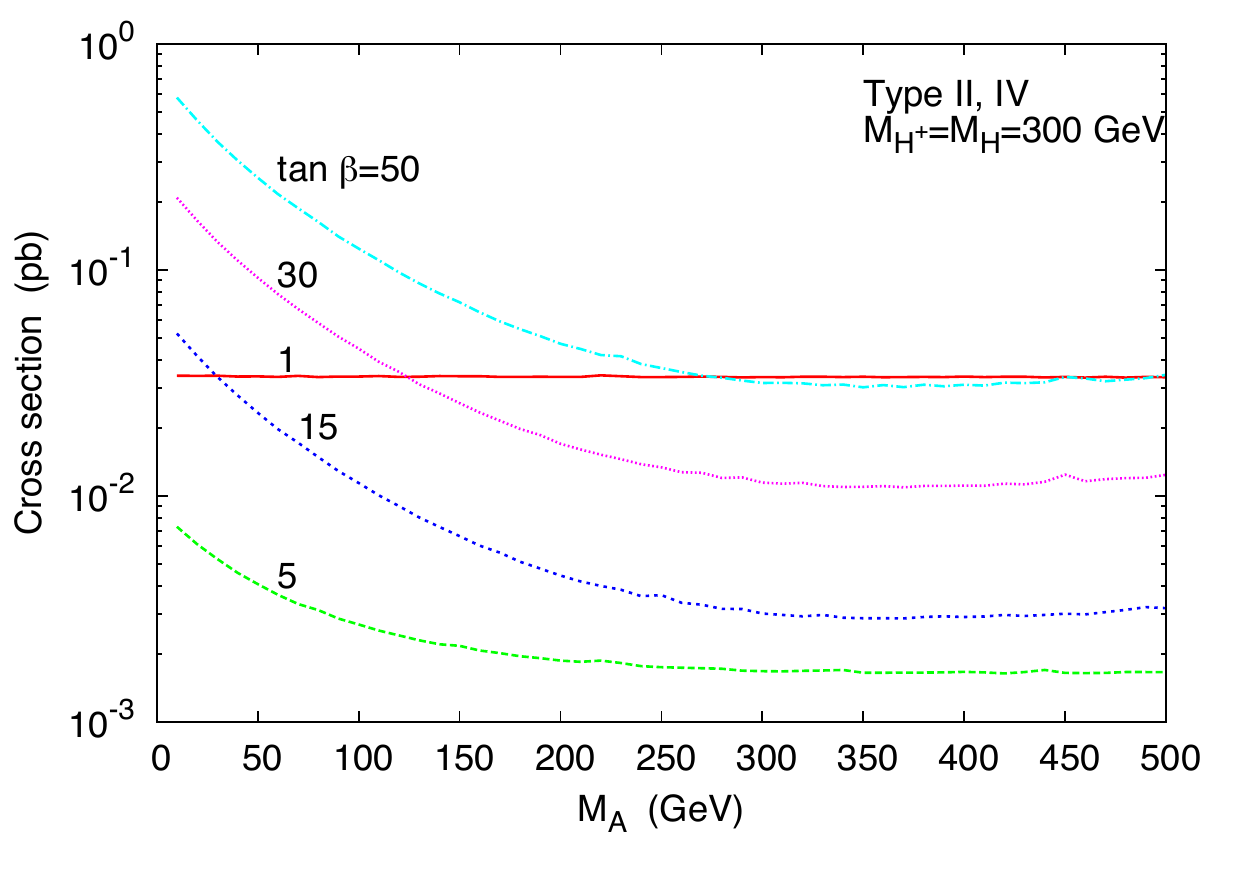}
\caption{\small \label{fig:cx2}
The $pp\to H^\pm b j$ cross sections at LHC-14 as a function of $M_A$ 
in the 2HDM type I and III (left), and II and IV 
(right) for several values of $\tan\beta$. We have taken
$M_{H^\pm}=M_H=300$ GeV. }
\end{figure}

It is well known that in the MSSM and for $M_A\geq 200$ GeV
all the heavy Higgs bosons become degenerate $M_H=M_A=M_{H^\pm}$ and
$\cos(\beta-\alpha)\to 0$. In general 2HDMs all Higgs boson masses are 
independent parameters. One can then identify the lightest CP-even with 
the observed 125 GeV Higgs boson and take the others  
$M_H$, $M_A$ and $M_{H^\pm}$  as free parameters.
In Fig.~\ref{fig:cx2}, we show the total cross sections as a 
function of the CP-odd Higgs mass for a few values of 
$\tan\beta=1-50$ and $M_H=M_{H^\pm}=300$ GeV in 2HDMs types I and III
(left panel) and types II and IV (right panels).
Note that for this choice of masses the production cross section
is dominated by the $\bar b$-initiated process.
For the values of the couplings in production cross sections, we refer to
Table~\ref{tab:1to4}.
In types I and III, the cross sections are insensitive to the CP-odd 
Higgs mass and they are
suppressed by $1/\tan^2\beta$ with increasing $\tan\beta$. 
The largest value of the cross section is obtained at $\tan\beta=1$ and is 
of the order $27$ fb. 
In types II and IV, one can see some sensitivity to the CP-odd Higgs mass. 
For $M_A \leq 250$ GeV, the cross section increases for lighter CP-odd 
Higgs mass and becomes almost constant for $M_A\geq 250$ GeV.
The enhancement of the cross section for $M_A\leq 250$ GeV is amplified with
large values of $\tan\beta$. For $M_A=100$ GeV and $\tan\beta=30$ one can
reach a cross section of the order of 40 fb.

\begin{figure}[t!]
\includegraphics[width=5.2in]{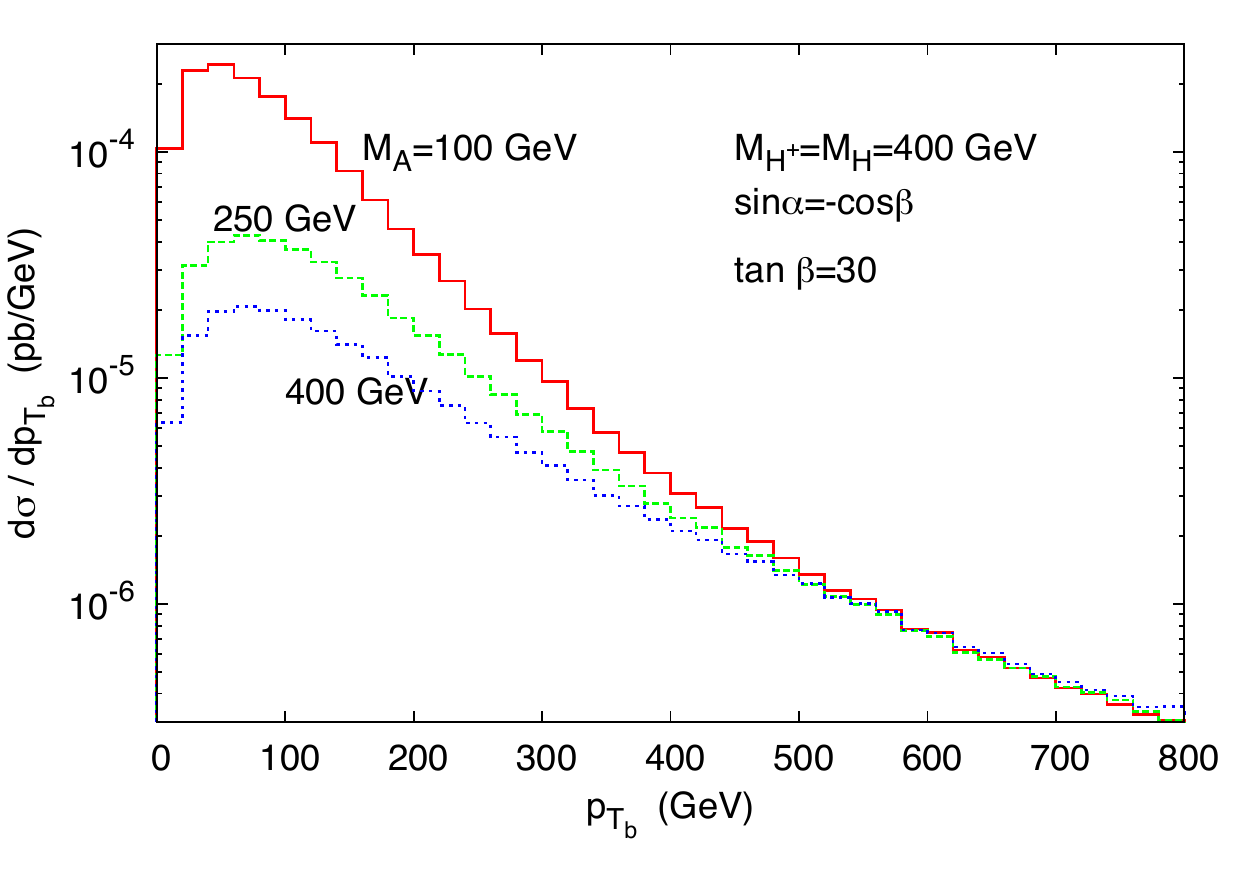}
\caption{\small \label{fig:ptb}
The transverse momentum $p_{T_b}$ distribution 
for $pp\to H^\pm b j$ in 2HDM type II and IV at the LHC-14 for
$M_A = 100, 250, 400$ GeV.
We have taken the decoupling limit $\sin\alpha=-\cos\beta$, $\tan\beta=30$,
and $M_{H^\pm}=M_H=400$ GeV.}
\end{figure}

In Fig.~\ref{fig:ptb}, we plot the $p_{T_b}$ distribution 
of the $b$ quark in the decoupling limit and for $M_{H^\pm}=M_H=400$ GeV
and $\tan\beta=30$ for several values of $M_A=100, 250$ and $400$ GeV.
As one can see from the plot, the distribution is enhanced for light 
$M_A=100$ GeV and for $p_{T_b}\leq 200$ GeV.
The transverse momentum of the $b$ quark is then a useful variable to
separate the contributions between the top diagram and the $W$-$A$ 
fusion diagram. One can require $p_{T_b} < 200$ GeV to suppress the
top-exchange contribution.
Therefore, we can see that the $W$-$A$ fusion diagram dominates for 
light $M_A$ and at the lower $p_{T_b}$ region.

\subsection{Large $\tan\beta$ and LHC $pp\to \Phi \to \tau^+ \tau^-$ data}
\begin{figure}[t!]
\includegraphics[width=3.2in]{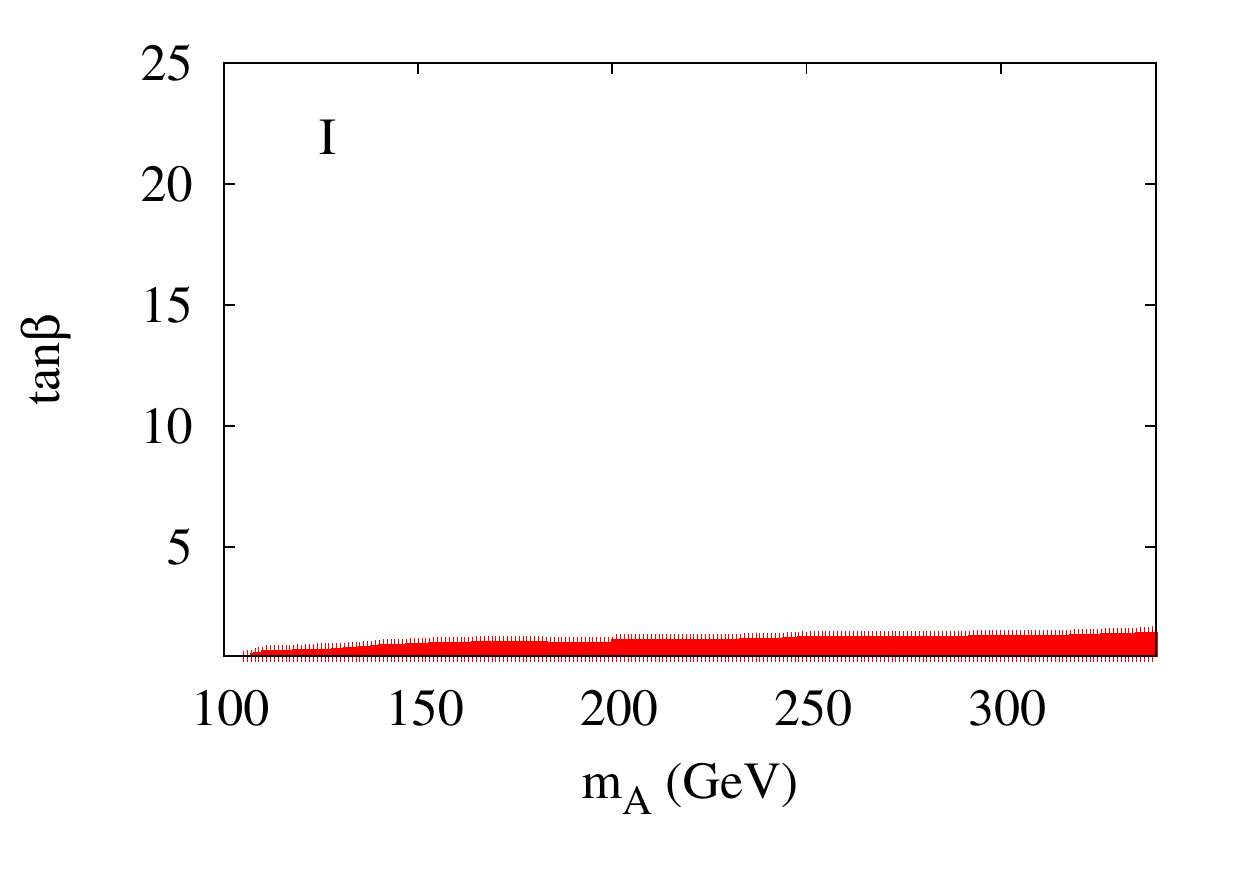} 
\includegraphics[width=3.2in]{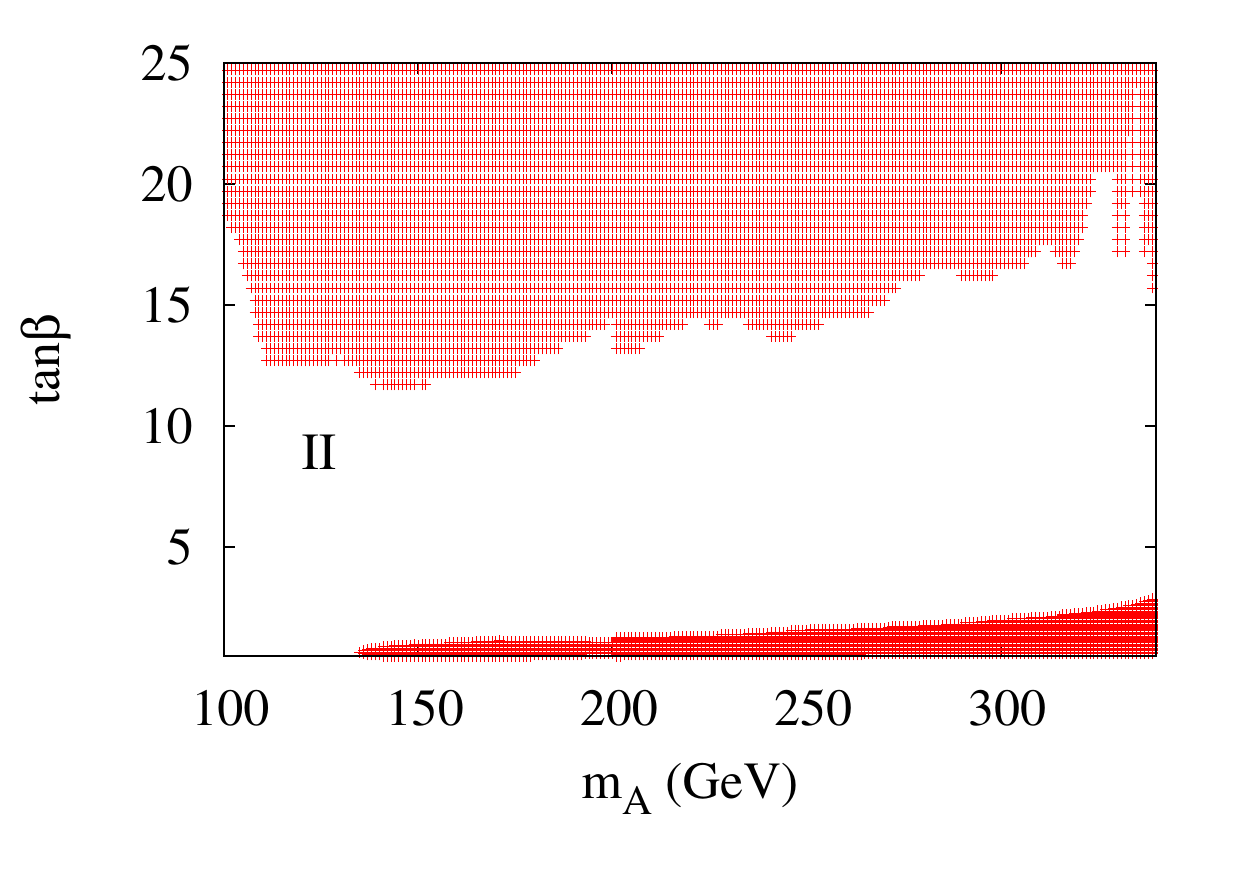} 
\includegraphics[width=3.2in]{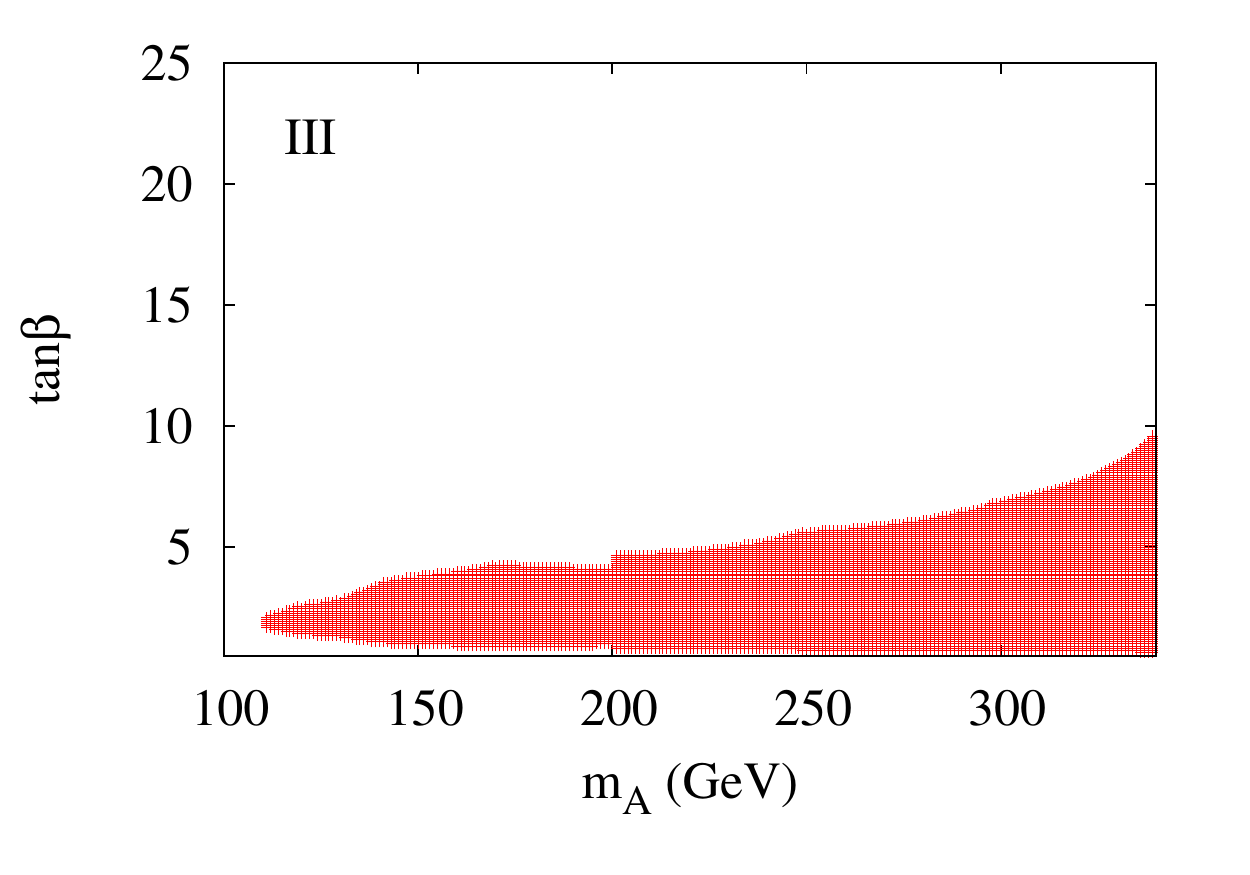} 
\includegraphics[width=3.2in]{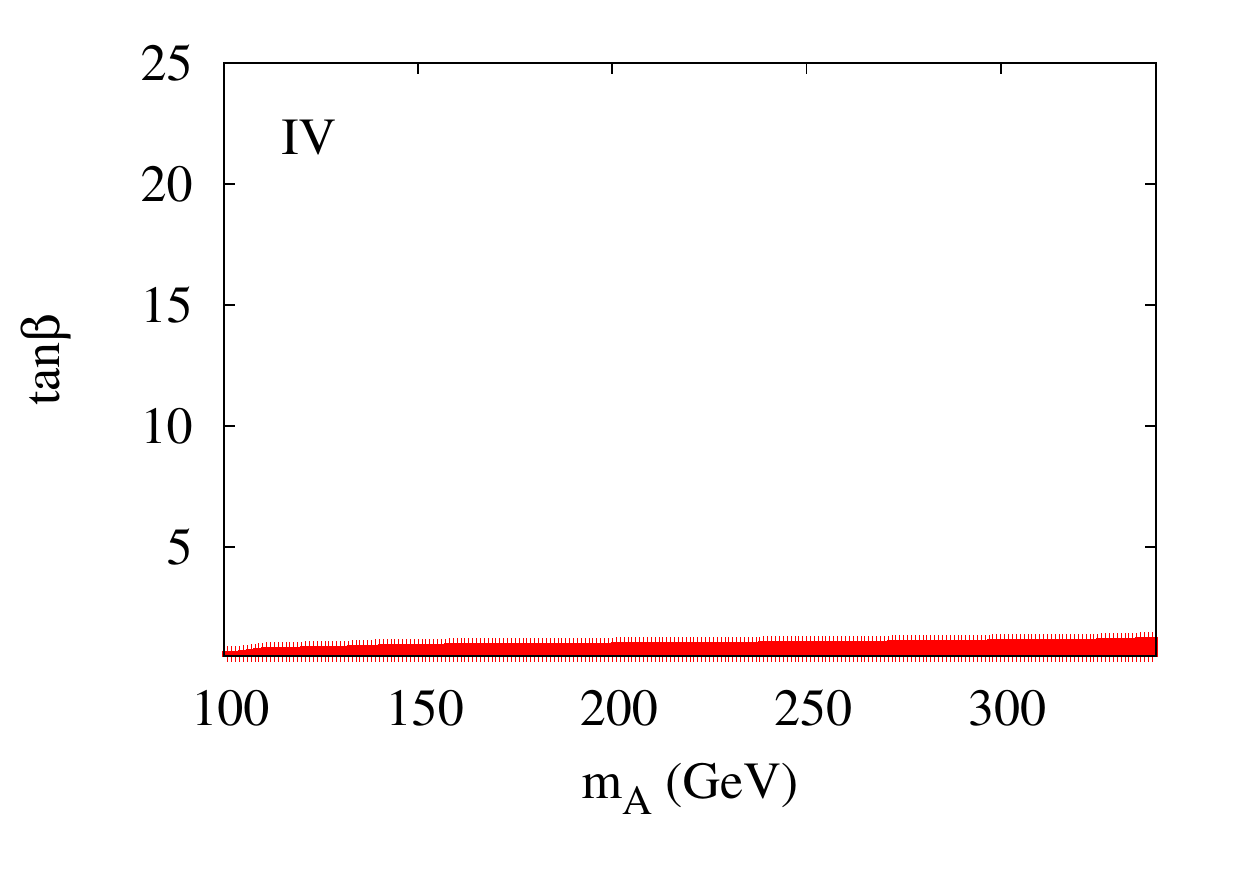} 
\caption{\small \label{fig:excl}
Exclusion plots in the plane of $(M_A,\tan\beta)$ 
for the 2HDMs type I, II, III and IV, using 
the ATLAS data for $gg\to A \to \tau^+ \tau^-$. 
The excluded region is shown in red while the rest is allowed.}
\end{figure}

At the LHC with 7 and 8 TeV, searches for the Higgs bosons $\Phi$ that 
decay into tau pairs, which in turn decay into those final states 
with one or two light leptons, have been performed 
\cite{Aad:2014vgg,CMS:2013hja} for Higgs mass in the range $[100,900]$ GeV.
Both ATLAS and CMS have some exclusion limits 
given as $\sigma(gg\to \Phi) \times B(\Phi \to \tau^+ \tau^-)$ as a function
of the Higgs mass $m_\Phi$. 
These limits can be interpreted in the 2HDMs if we take the 
Higgs state $\Phi$ as one of the neutral Higgs bosons of the 2HDMs: 
$\Phi=h$ and/or $H, A$. 
In fact if $h$ mimics the SM Higgs boson, $\sigma(gg\to h) \times 
B(h \to \tau^+ \tau^-)$ will not have any enhancement factor such as 
$\tan\beta$. 
 Since
we have observed that 
in order to enhance $pp\to b H^+ j$ cross sections one needs both 
non-degenerate $A$ and $H$ and also large $\tan\beta$, 
here we attempt to find 
what would be the largest possible value 
for $\tan\beta$ such that it is still consistent 
with $\tau^+ \tau^-$ data for $100 \leq M_A \leq 340$ GeV and 
assuming that the heavy CP-even Higgs boson is rather heavy. 
A similar study with the 7 TeV data had been done 
in \cite{Arhrib:2011wc} for 2HDM.
Because of CP invariance the CP-odd Higgs boson $A$ does not couple to $WW$
or $ZZ$, and the partial decay widths into loop mediated 
$gg$, $\gamma\gamma$, and $\gamma Z$ channels
are highly suppressed. 
The decay channel $A\to  hZ$, which is proportional to
$\cos(\beta-\alpha)$, will also be severely suppressed if we assume that 
$\beta-\alpha$ is close to the decoupling limit. 
Therefore, the CP-odd Higgs boson predominantly decays into fermion pairs: 
$q\bar{q}$, $q=b,s,d,c,u$ and $l^+l^-$ $l=\tau, \mu, e$. 

In 2HDM-I and -IV, the coupling $A\tau^+\tau^-$ is proportional to 
$1/\tan\beta$  while in 2HDM-II and -III it is proportional to $\tan\beta$.
On the other hand, from Table~\ref{tab:2hdtype}
the coupling of $Ab\bar b$ is proportional to
$\tan\beta$ in  2HDM-II and -IV but $1/\tan\beta$ in  2HDM-I and -III.
Thus, it is clear that in 2HDM-I (resp. II) both the 
production rate $gg\to A$ and the decay $A\to \tau^+\tau^-$ are 
suppressed (resp. enhanced) 
for large $\tan\beta$. 
We then expect a strong exclusion for large $\tan\beta$
in type II but not in type I. 
In all four 2HDM types we expect some
enhancement for small $0.5 \leq \tan\beta \leq 1$ because $At\bar{t}$
is proportional to $m_t/\tan\beta$.

For a given CP-odd Higgs mass $M_A$ and $\tan\beta$, 
the cross section of $gg\to A$, which only depends on these 2 parameters, 
is computed with help of SUSHI public code \cite{Harlander:2012pb}. 
In the decoupling limit  $A\to Zh$ is vanishing, and if 
$A\to \{ W^\pm H^\mp, ZH, t\bar{t}\}$
are closed, the branching ratio of $A\to \tau^+\tau$ depends on
$\tan\beta$ and $M_A$ only, and there is no $\sin\alpha$ dependence. 
Therefore, the cross section $gg\to A\to \tau^+ \tau^-$ 
will depend only on  $\tan\beta$ and $M_A$. Hence, our exclusion from 
$\tau^+\tau^-$ data can be given in the plane of $(M_A,\tan\beta$). 
After computing the cross section $gg\to A$ times the branching fraction
of $A\to \tau^+\tau^-$, we compare our theoretical predictions
with the ATLAS data \cite{Aad:2014vgg}. 
Note that the ATLAS data were given for 
$M_A=90, 100, ... 340$ GeV with steps of 10 GeV or even larger in some cases.
Therefore, we have used linear interpolations for $M_A$ values in-between 
the data.

We draw in Fig.~\ref{fig:excl} the exclusion region in $(M_A, \tan\beta)$
plane for 2HDMs type I, II, III, and IV,  where $M_A$ is in the 
range $[100,340]$ GeV and $0.5\leq \tan\beta\leq 50$.
In 2HDM types I and IV, there is no exclusion for $\tan\beta\geq 1.5$.
The reason is that in type I, the production rate of $gg\to A$ and 
decay of $A\to \tau^+ \tau^-$  are both suppressed by
$1/\tan\beta$. 
Whereas in 2HDM type IV, the production rate is enhanced by the bottom
Yukawa for large $\tan\beta$ but the decay $A\to \tau^+\tau^-$ is suppressed
by $1/\tan\beta$ which cancels the bottom enhancement in the production,
which then gives no exclusion for large $\tan\beta$.
In 2HDM type III, similar to type I the production rate is 
suppressed by $1/\tan\beta$.
However, the branching fractions of $A\to q\bar{q}$ are suppressed by 
$1/\tan\beta$ while $B(A\to \tau^+ \tau^+)$ is enhanced for large
$\tan\beta$. For this reason the exclusion in type III is somewhat
 stronger than that in type I. There is no exclusion for 
$\tan\beta\geq 9.5$ (resp. $\tan\beta\geq 2$) for $M_A\approx 340$ GeV
(resp. for $M_A\approx 112$ GeV). 
On the other hand, the 2HDM-II receives both enhancement at 
large $\tan\beta$ for  the production rate $gg\to A$ and 
the $B(A\to \tau^+\tau^-)$. Thus, this gives a strong exclusion for 
$\tan\beta \geq 22$ for all $M_A\in [100,340]$ GeV. 
The low $\tan\beta \leq 3$ region (resp. $\tan\beta \leq 1$) is also excluded 
for $M_A=340$ GeV (resp. $M_A=150$ GeV).

\section{Beyond Two-Higgs-Doublet Models}
\label{sec:b2hdm}
Another interesting possibility to enhance the production of charged Higgs 
boson might be the case in which a 2HDM is not an
ultraviolet (UV) complete theory and the UV cutoff locates far above the 
mass scale of heavy Higgs bosons
\footnote{
%
We note that this kind of enhancement arising from unitarity violation 
might be dangerous and should be taken with caution, 
because a UV-complete model might contain 
new interactions to restore the unitarity. 
}.
In this case,
taking one of the 2HDMs as a low-energy reference model,
the relevant interactions may be parameterized as
\begin{eqnarray}
{\cal L}_{H_i\bar{b}b}&=&-\frac{gm_b}{2m_W}\,
\bar{b}\left(\xi^S_i\,g^S_i+i\,
\xi^P_i\,g^P_i\,\gamma_5\right)b\,H_i \,, \nonumber \\[2mm]
{\cal L}_{H^\pm tb}&=&+\frac{gm_b}{\sqrt{2}m_W}\,
\bar{b}\left(\xi_L\,c_L\,P_L+ \xi_R\,c_R\,P_R\right)
t\,H^- \ + \ {\rm h.c.}\,, \nonumber \\[2mm]
{\cal L}_{H_iH^\pm W^\pm}&=&-\frac{g}{2}\,(\xi_S\,S_i + i\xi_P\,P_i)\,
\left[H^-\left(i\stackrel{\leftrightarrow}
   {\partial_\mu}\right)H_i\right]\,W^{+\mu}
\ + \ {\rm h.c.}\,.
\end{eqnarray}
In the MSSM, for example, including the $\tan\beta$-enhanced 
SUSY threshold corrections to the  down-type Yukawa couplings, we have
\begin{equation}
\label{mssm1}
\xi^S_i=\xi^P_i=\xi_L=\frac{1}{1+\kappa_b\tan\beta}\,, \ \ \
\xi_R=\xi_S=\xi_P=1\,,
\end{equation}
with~\cite{Lee:2004}
\[
\kappa_b
= \epsilon_g + \epsilon_H ,
\]
where $\epsilon_g$ and $\epsilon_H$ are the contributions from the
sbottom-gluino exchange diagram and from stop-Higgsino diagram,
respectively. Their explicit expressions are
\[
\epsilon_g = \frac{2\alpha_s}{3\pi}M^*_3\mu^*
I(m^2_{\tilde{b}_1},m^2_{\tilde{b}_2},\vert M_3\vert^2) , \qquad
\epsilon_H = \frac{\vert h_t\vert^2}{16\pi^2}A^*_t\mu^*
I(m^2_{\tilde{t}_1},m^2_{\tilde{t}_2},\vert \mu \vert^2)\ ,
\]
where $M_3$ is the gluino mass parameter, $h_t$ and $A_t$ are the
top-quark Yukawa and trilinear couplings, respectively.

\begin{figure}[t!]
\includegraphics[width=3.2in]{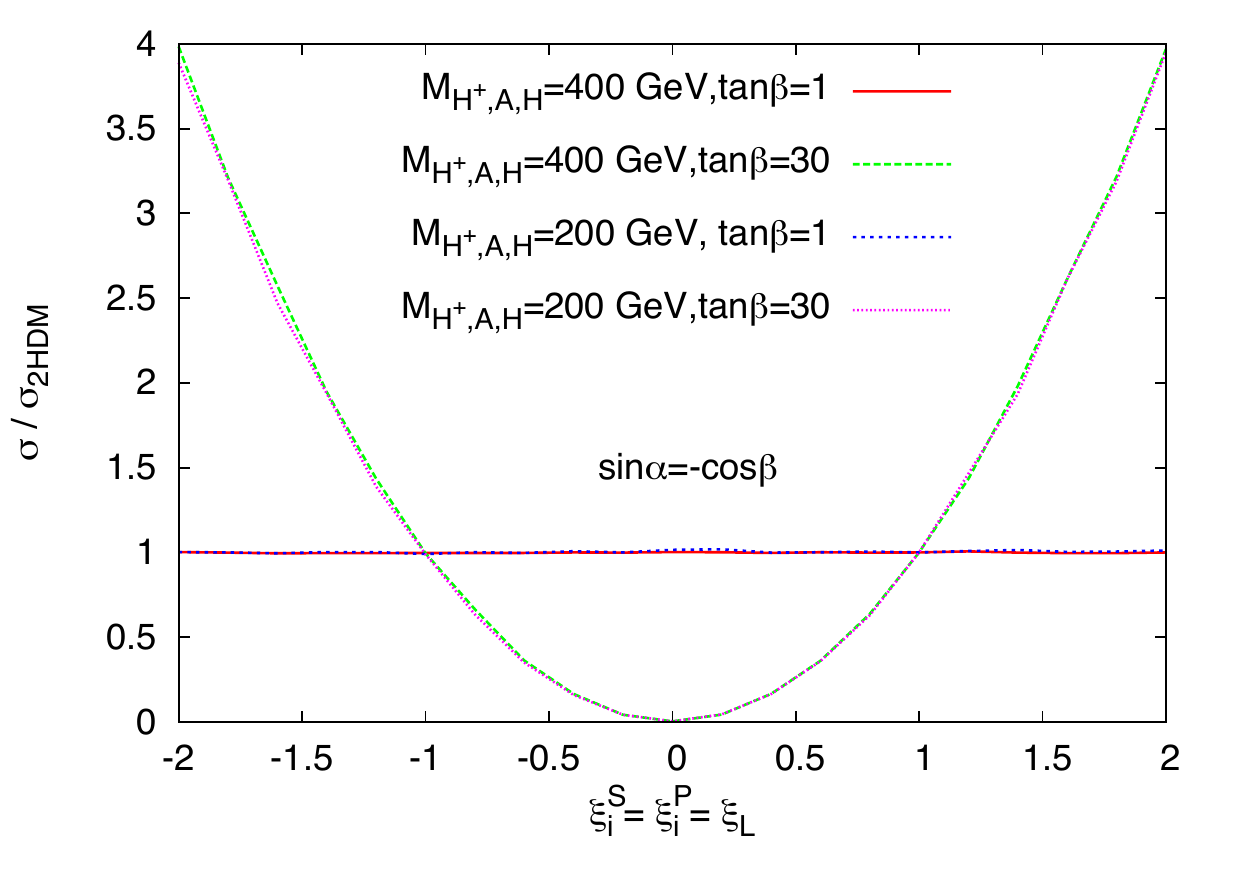}
\caption{\small \label{fig:ratio1}
Ratios of cross sections for varying $\xi_i^S = \xi_i^P = \xi_L$ 
to the cross section at the 2HDM type II value ($\xi_i^S = \xi_i^P = \xi_L=1$),
showing the enhancement due to the modification of the 2HDM as suggested
in the text.
}
\end{figure}

\begin{figure}[t!]
\includegraphics[width=3.2in]{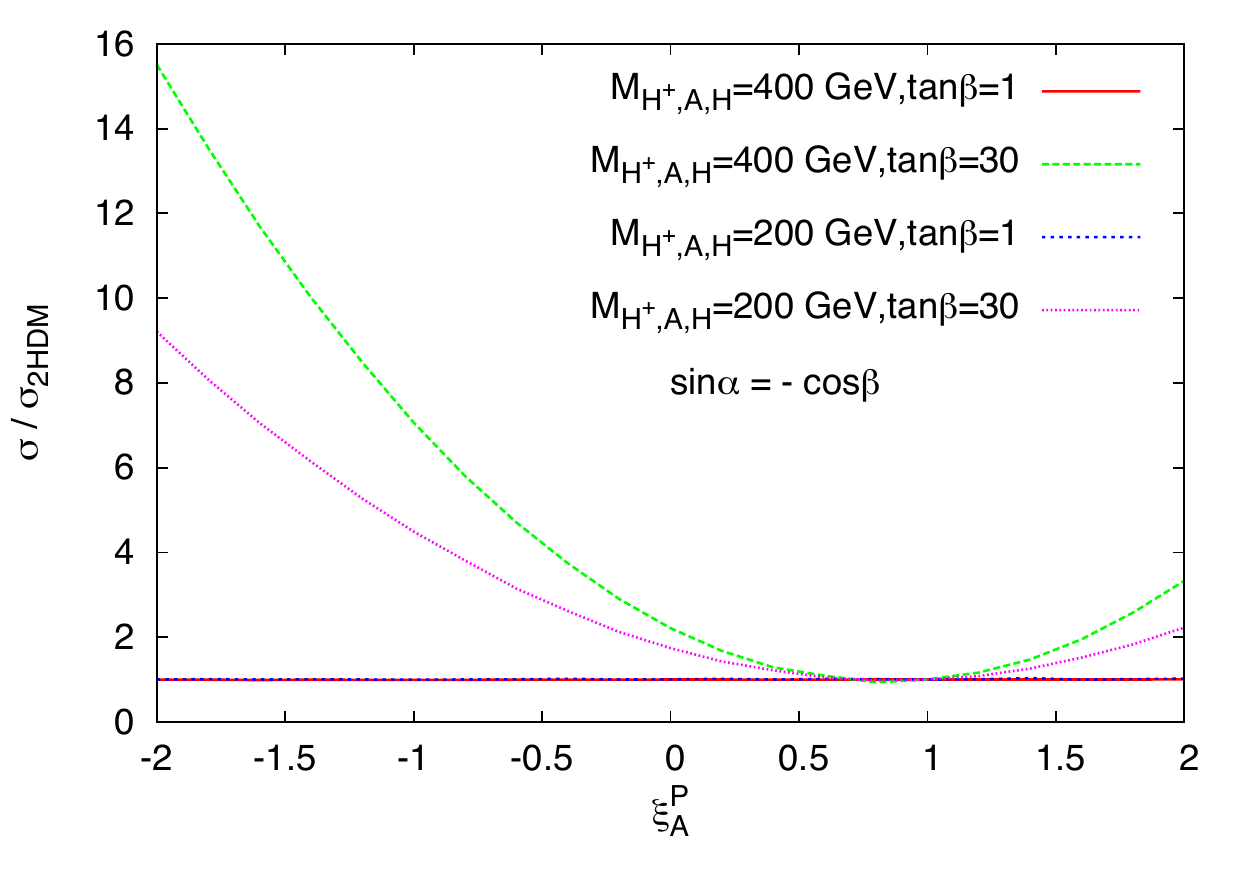}
\includegraphics[width=3.2in]{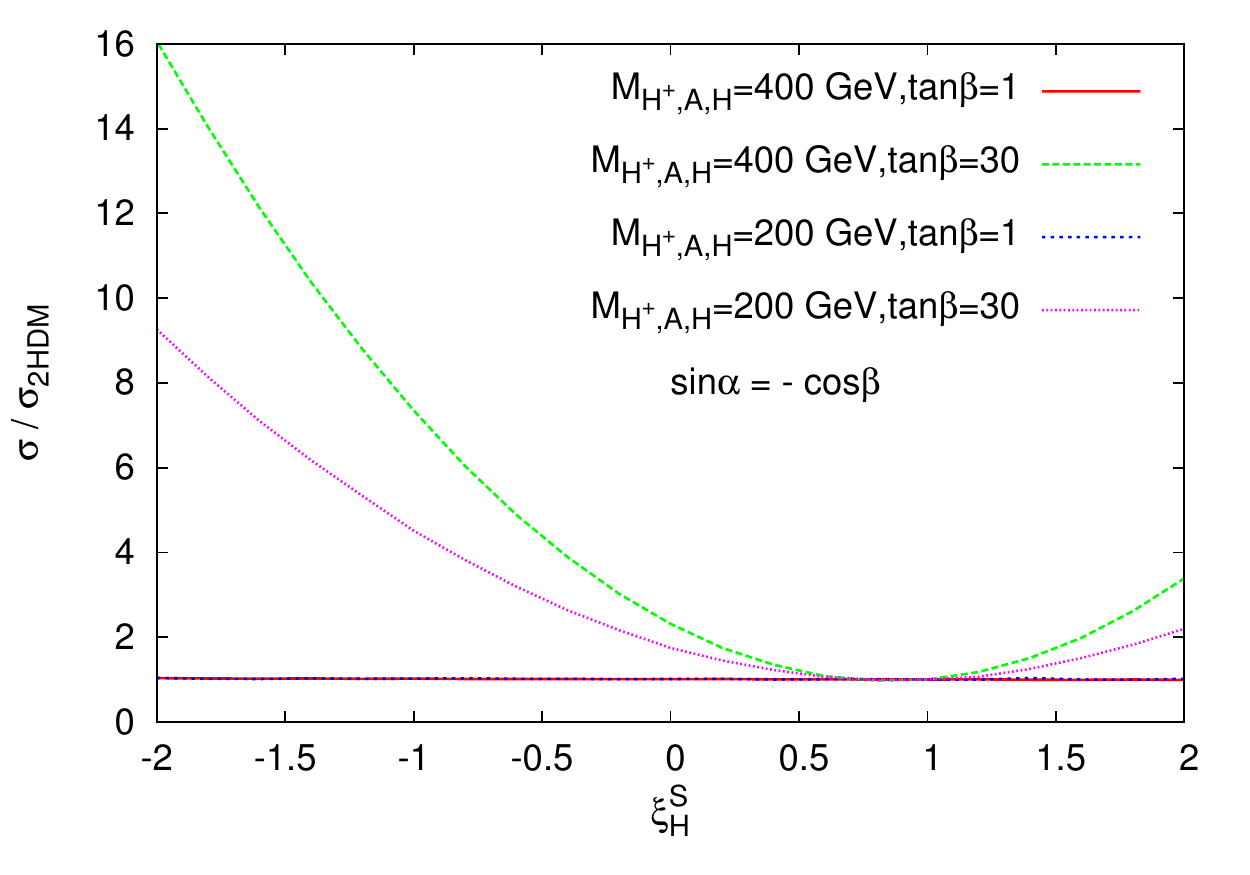}
\caption{\small \label{fig:ratio2}
Ratios of cross section with varying $\xi_A^P$ (left panel) or 
$\xi_H^S$ (right panel) to the cross section at the 2HDM value 
(i.e. $\xi_A^P=1$ or $\xi_H^S=1$ respectively) by keeping all other
parameters at their 2HDM type II values.
}
\end{figure}

Without loss of generality we choose the 2HDM type II as the 
reference model, and show the change in production cross sections
with the variations in the couplings $\xi_i^{S,P}$ and $\xi_L$. We used
Eq.~(\ref{mssm1}) as the guidance. We first show the ratio of the
cross sections for varying $\xi_i^S = \xi_i^P = \xi_L$ between $-2$ and $+2$
to the cross section at the 2HDM type II values, i.e., 
$\xi_i^S = \xi_i^P = \xi_L=1$ in Fig.~\ref{fig:ratio1},
for $\tan\beta = 1,30$.
The $\tan\beta=1$ curves show almost no sensitivity
to $\xi_i^S = \xi_i^P = \xi_L$ because the process is dominated
by the top-Yukawa term.
On the other hand, the $\tan\beta=30$ curves are dominated by the
bottom-Yukawa term.
It is obvious that the ratio is close to zero for $\xi_i^S = \xi_i^P = \xi_L=0$,
and is one for $\xi_i^S = \xi_i^P = \xi_L=1$. The ratio grows as the square
of the couplings around 0 to almost 4 at 
$\xi_i^S = \xi_i^P = \xi_L = \pm 2$.

If for some higher scale dynamics such that $\xi^S_i$ and $\xi^P_i$ do not
change in the same manner, we show the effects on the cross sections
in Fig.~\ref{fig:ratio2}. On the left panel, we show the ratio of 
cross sections with varying $\xi_A^P$ between $-2$ and $+2$ to the 
cross section at the 2HDM value (i.e. $\xi_A^P=1$) by keeping all other
parameters at their 2HDM type II values. Again, the sensitivity at
$\tan\beta=1$ is negligible, while it becomes quite nontrivial for
large $\tan\beta=30$. As we have shown in Sec. III that the $W$-$A$ 
diagram interferes destructively with the top diagram, we can now
turn the destructive interference into constructive one by reversing 
the sign of $\xi_A^P$.
Furthermore, when $\xi_A^P$ is negative the second term 
in Eq.~(\ref{22}) would not vanish. 
It is then very clear that the ratio becomes
quite large at negative $\xi_A^P$. At $\xi_A^P= 0$, the ratio is already
larger than 1 because no interference comes from the $W$-$A$ diagram.
Similar behavior occurs for $\xi_H^S$ as shown on the right panel
in Fig.~\ref{fig:ratio2}.  The ratios that $\xi_H^S$ can attain are
very similar to those of $\xi_A^P$.

\section{Conclusions}
\label{sec:con}

We have performed the study of
$b$- and $\bar b$-initiated processes of 
$pp \to j H^\pm b/\bar b$ in the 2HDM framework 
at the LHC-14 in the decoupling limit
$(\sin\alpha = - \cos\beta)$, which is favored by the current Higgs data.
We have identified strong cancellations between the top diagram and 
the $W$-$H_i (H_i=h,H,A)$ diagrams which rendered the process very 
suppressed.  The cancellation is indeed the strongest
when the $A$ and $H$ are degenerate and in the decoupling limit. 
We pointed out that if the pseudoscalar Higgs boson $A$ is much lighter 
than the CP-even Higgs boson $H$, the cross section of charged-Higgs
production can be substantially enhanced, because the cancellation
is no longer complete.  
We have explicitly obtained the exclusion in parameter space of
$(M_A,\tan\beta)$ for 2HDM types I to IV based on the LHC data on
$\sigma(g \to \Phi)\times B(\Phi \to \tau^+ \tau^-)$.
In the allowed paramete space, 
the size of production cross section
can be as large as ${\cal O}(50)$ fb for $M_A = 100$ GeV and $\tan\beta=30$
for types II and IV.
This is the main result of the work.

We offer the following comments on the findings of this work as follows.
\begin{enumerate}
\item The $b$-initiated process for production of $H^+$ in
$q b \to q H^+ b$ suffers from a very strong cancellation between
the top diagram and the $W$-$H_i$ diagrams. However, the
$\bar b$-initiated process for production of $H^+$ in
$q \bar b \to q H^+ \bar b$ suffers a less severe cancellation, mainly
because of the $u$-channel top-exchange instead of $s$-channel.

\item The strong cancellation is dictated by the absence of the
unitarity-breaking terms and we find the sum rules
expressed by
the relevant Higgs couplings, see Eq.~(\ref{eq:sumrules}).

\item
For $M_{H^\pm} \le m_t - m_b$ the top diagram completely dominates as the
top quark is produced on-shell as like that in single-top production.
However, when $M_{H^\pm} > m_t - m_b$ the $W$-$H_i$ fusion diagrams also
contribute.

\item
In the future study, we shall make use of the special kinematics, e.g,
the $p_{T_b}$ distribution, to discriminate between the top and 
the $W$-$H_i$ fusion diagrams. The goal is to isolate the effect
of light pseudoscalar Higgs boson, which is still allowed by the current data.

\item
Current LHC data on $\sigma (gg \to \Phi)\times B(\Phi \to \tau^+\tau^-)$
constrains the parameters of the 2HDMs. We found that the data
constrains the most on type II because both the production process 
and the decay are $\tan\beta$ enhanced. Yet, there are sizable allowed
parameter space between $\tan\beta =3$ and $22$. 
The second most constrained is type III because the production rate
is suppressed by $1/\tan\beta$ but the decay is enhanced by $\tan\beta$. 
Types I and IV have the most available parameter space. 

\item 
The process $pp \to j H^\pm b/\bar b$ that we consider in this work would be
more interesting for type II and IV because of larger cross sections. 
Especially type IV has the least restriction from the current
LHC data on $\sigma (gg \to \Phi)\times B(\Phi \to \tau^+\tau^-)$,
and it can allow cross sections as large as $O(100-300)$ fb for
$\tan\beta = 30-50$ and $M_A=50-100$ GeV.

\item 
When the 2HDM is a low-energy limit of some ultraviolet (UV) models, 
the integrity of the Yukawa couplings may change. For example, 
in the MSSM the SUSY 
particles can largely change the bottom-Yukawa couplings with strong
and weak interactions. Varying the bottom Yukawa coupling of either
$A$ or $H$ gives non-trivial behavior for the production cross sections.

\end{enumerate}

\section*{Acknowledgment}  
We thanks Marco O. P Sampaio for providing us with CP-odd cross sections.  
K.C. was supported by the MoST of Taiwan under Grants number 
102-2112-M-007-015-MY3.  
J.S.L. was supported by
the National Research Foundation of Korea (NRF) grant
(No. 2013R1A2A2A01015406).
A.A would like to thank NCTS for warm hospitality 
where part of this work has been done.

\end{document}